\documentclass[10pt,twocolumn,twoside]{IEEEtran}
\usepackage{cite}

\ifCLASSINFOpdf
   \usepackage[pdftex]{graphicx}
\else
\fi
\usepackage{amsmath}
\usepackage{amssymb}
\usepackage{mathrsfs} 
\usepackage{algorithmic}

\newtheorem{theorem}{Theorem}
\newtheorem{corollary}{Corollary}[theorem]
\newtheorem{assumption}{Assumption}

\ifCLASSOPTIONcompsoc
  \usepackage[caption=false,font=normalsize,labelfont=sf,textfont=sf]{subfig}
\else
  \usepackage[caption=false,font=footnotesize]{subfig}
\fi
\begin{document}
%
\title{Adaptive Online Distributed Optimal Control of Very-Large-Scale Robotic Systems}
%
%
%

\author{Pingping~Zhu,~\IEEEmembership{Member,~IEEE,}
        Chang~Liu,~\IEEEmembership{Member,~IEEE,}
        and~Silvia~Ferrari,~\IEEEmembership{Senior~Member,~IEEE}
\thanks{P. Zhu, C. Liu, and S. Ferrari are with the Sibley School of Mechanical
	and Aerospace Engineering, Cornell University, Ithaca, NY 14853 USA (e-mail: pingping.zhu@cornell.edu; cl775@cornell.edu; sferrari@cornell.edu)}
}

\maketitle

\begin{abstract}
This paper presents an adaptive online distributed optimal control approach that is applicable to optimal planning for very-large-scale robotics systems in highly uncertain environments. This approach is developed based on the optimal mass transport theory. It is also viewed as an online reinforcement learning and approximate dynamic programming approach in the Wasserstein-GMM space, where a novel value functional is defined based on the probability density functions of robots and the time-varying obstacle map functions describing the changing environmental information. The proposed approach is demonstrated on the path planning problem of very-large-scale robotic systems where the approximated layout of obstacles in the workspace is incrementally updated by the observations of robots, and compared with some existing state-of-the-art approaches. The numerical simulation results show that the proposed approach outperforms these approaches in aspects of the average traveling distance and the energy cost.

\end{abstract}


%
\IEEEpeerreviewmaketitle

\section{Introduction}
%
%
%
%
\IEEEPARstart{W}{ith} the development of low-cost sensor, wireless communication, and embedded computational systems, very-large-scale robotic (VLSR) systems comprised of hundreds of autonomous
robots are becoming a promising research area, where a large number of mobile robots can cooperate to complete a given task more efficiently and effectively than a single or a few mobile robots. In the past two decades, significant progress of VLSR systems has been made in different research methodologies, including the optimal control (OC) \cite{StengelOptimalControl1994,ParkerPathPlanningMotionCoordination2009,RuddIndirectDOC2013,FoderaroDOC2014,FoderaroDOC2016,FerrariDOCtutorial2016,RuddIndirectDOC2017} and multi-agent reinforcement learning (MARL) \cite{YangMeanFieldMARL2018,KhanScalableCentralizedDMARL2018,HuttenrauchDeepRLforSR2019,HernandezSurveyMultiagentDRL2019,CarmonaLinearQuadraticMeanField2019,CarmonaModelFreeMeanFieldRL2019,ZhangReviewMARL2019,RenRepresentedVFforLSMARL2020}. Some similar or identical tasks and applications of VLSR systems are considered and solved from different points of view, including the multi-agent path planning \cite{FerrariMultirobotMotionCoordination1998,BennewitzDecoupledPathPlanning2002} and the control of swarm robotics (SR) \cite{YogeswaranSwarmRoboticsReview2010,RubensteinProgrammableSelfAssembly2014,BandyopadhyayProbabilisticSwarmGuidanceOT2014,KrishnanDistributedOTforSwarm2018,BayindirSwarmRoboticsReview2016,NedjahSwarmRoboticsReview2019}.

One significant challenge of VLSR systems is the scalability issue, which is also referred to as the combinatorial nature of MARL in \cite{HernandezSurveyMultiagentDRL2019}. Even in a given and known environment, the optimization of the plans of $N$ cooperative robots is PSPACE-hard \cite{ParkerPathPlanningMotionCoordination2009}, which is not acceptable for a VLSR system with a very large $N$. Thus, several approaches are proposed to represent the macroscopic state of the VLSR system instead of the microscopic state of every individual robot, including the Nash Certainty Equivalence (NCE) or mean field, \cite{HuangNCE2006,HuangNEC2010}, and the distributed optimal control (DOC) \cite{RuddIndirectDOC2013,FoderaroDOC2014,FoderaroDOC2016,FerrariDOCtutorial2016,RuddIndirectDOC2017} approaches.  In the NCE approach, the macroscopic state is the mass of all robots, and the key idea is to specify a certain consistency relationship between the individual robot kinodynamics and the mass effect, where the mean field coupling is produced by the averaging of the microscopic robot kinodynamics and costs. While, in the DOC approach, the macroscopic state is the time-varying probability density function (PDF) of robots and the cost does not depend on the microscopic state of every individual robot, and, thus, the coupling is not needed. Several variants of the DOC approach have been developed to solve the path planning problem of VLSR systems \cite{RuddIndirectDOC2013,FoderaroDOC2014,FoderaroDOC2016,FerrariDOCtutorial2016,RuddIndirectDOC2017}, where the optimal trajectory of the robot PDFs from a given initial distribution to a given target distribution is generated as a reference via nonlinear programming (NLP) to guide these robots to travel in the region of interest (ROI) with fixed and known  obstacles. 

Another significant challenge of the VLSR systems is the adaptability issue. In many emerging systems, the goal is to control multiple assets and resources in the presence of significant uncertainties that cannot be modeled a priori. The VLSR systems are required to respond to significant changes in the environment and target information that occur over long time scales and re-plan and adapt to local uncertainties while preventing network-level instabilities. Although the DOC approach has been shown to overcome the computational complexity associated with classical optimal control by representing the state of the VLSR systems by the robot PDFs, 
its computational complexity is relatively high because of the NLP on the continuous spaces of robot microscopic state and control. 
It is not feasible for the DOC approaches to re-plan the optimal trajectory of robot PDFs in real time. Thus, existing DOC approaches do not satisfy the requirements in the applications involved in the  uncertain and changing environments. Recently, a model-free MARL approach was proposed based on mean field control (MFC) \cite{CarmonaModelFreeMeanFieldRL2019}, where the macroscopic state of the VLSR system is also the robot PDF and the rewards or Lagrangian terms depend on both microscopic and macroscopic states. This MARL approach can be recast as a Markov decision process (MDP) on the Wasserstein space of measures and implemented by learning a  deterministic control law offline. The control law is a functional with the arguments of the macroscopic and microscopic states. However, the value functional and the corresponding control law cannot be approximated efficiently, especially for the VLSR systems with continuous states and controls. Thus, this MARL approach cannot be applied in the uncertain and changing environments neither.       

In this paper, an adaptive DOC (ADOC) theory and approach for VLSR systems are proposed to carry out online cooperative sensing and navigation tasks in highly uncertain environments. Since the similarities between the optimal control and reinforcement learning, the proposed ADOC approach can be treated as a reinforcement learning-approximate dynamic programming (RL-ADP) approach \cite{LewisRLAPD2013} for which the environment has to be explored online.

Compared to existing works, this paper makes the following contributions: 1) The time-varying  environmental information is expressed by a map function and reflected in the Lagrangian term (or reward). 2) A new value functional is defined based on the time-varying environmental information. 3) The ADOC approach is formulated in a Wasserstein-GMM space based on the optimal mass transport (OMT) theorem where the robot PDFs are all assumed to be Gaussian mixture distributions. 4) The ADOC approach is implemented online via solving linear programming (LP) problems in a subspace of the Wasserstein-GMM space.  5) The effectiveness of the proposed ADOC approach is demonstrated on the problem of VLSR systems path planning with uncertain environmental information, and the results show that the ADOC approach outperforms the other three existing state-of-the-art approaches. 

\section{Problem Formulation}
\label{sec:Problem_Formulation}

Consider the problem of adaptively planning the trajectories of a VLSR system comprised of $N$ cooperative robots deployed according to a known distribution and tasked with forming a given target distribution through a large and obstacle-deployed region of interest (ROI), denoted by $\mathcal{W}\subset \mathbb{R}^2$. In the ROI, there are $N_B$ obstacles, $\mathcal{B}_1,\ldots,\mathcal{B}_{N_B} \subset\mathcal{W}$. The location, the geometry, and the number of obstacles are unknown a priori.  The actual layout of obstacles is the union of all obstacles denoted by $\mathcal{B} = \cup_{n_b = 1}^{N_{B}}\mathcal{B}_{n_b}$, and is unknown a priori as well. However, an approximate map is available for the VLSR system to indicate the layout of obstacles in the ROI denoted by $\hat{\mathcal{B}}_0 = \hat{\mathcal{B}}(t_0)\subset \mathcal{W}$ at the initial time $t_0$.  The approximate layout of obstacles $\hat{\mathcal{B}}(t) \subset \mathcal{W}$ can be updated by observations from all robots. The approximate layout of obstacles is represented by a time-varying function, $m(\mathbf{x},t)$ defined on  $\mathcal{W} \times \mathbb{R}$, referred to as obstacle map function. Let  $\mathscr{M}$ be a collection of all functions $m(\cdot,t)$ representing all possible layouts of obstacles in the ROI, such that $m(\cdot,t)\in \mathscr{M}$.


The dynamics of each robot are governed by a stochastic
differential equation (SDE),
\begin{align}
\label{eq:dynamics}
\dot{\mathbf{x}}_n(t) &= \mathbf{f}[\mathbf{x}_n(t), \mathbf{u}_n(t),t] + \mathbf{w}(t),\\
\mathbf{x}_n(t_0) &= \mathbf{x}_{n_0}, \hspace{10pt} n=1,...,N,
\end{align}
where $\mathbf{x}_n(t) \in \mathcal{W}$ denotes the $n$th robot's configuration at time $t$,
$\mathbf{u}_n(t) \in \mathcal{U}$ denotes the $n$th robot action or control, and
$\mathbf{x}_{n_0}$ denotes the initial configuration of the $n$th robot at the initial time
$t_0$. The robot dynamics in (\ref{eq:dynamics}) are characterized by an
additive Gaussian white noise, denoted by $\mathbf{w}(t) \in \mathbb{R}^2$. In this paper, for simplicity, assume that the position of the $n$th robot, $\mathbf{x}_n(t)$, can be measured accurately by an equipped GPS. 

To update the layout of obstacles, all robots are equipped
with identical omnidirectional range sensors. The field of view (FOV) of the $n$th sensor denoted by $\mathcal{F}_n(t) \subset \mathcal{W}$ can be represented by a circle of radius $r$ around $\mathbf{x}_n(t)$. Then, the whole FOV of all robots at time $t$ is denoted by $\mathcal{F}(t)=\cup_{n=1}^N \mathcal{F}_n(t)$. Assume that all robots can share information at any time, the obstacle at the position $\mathbf{x}\in\mathcal{W}$ is observed and updated at time $t$ if and only if $\mathbf{x} \in \mathcal{F}(t)$.



Since the robot positions, $\mathbf{x}_n(t) \in \mathcal{W}$, $n=1,\ldots,N$, at time $t $, are time-varying continuous vectors, assume that there exists a time-varying continuous probability distribution associated with a probability density function (PDF), $\wp(\mathbf{x},t) \in \mathscr{P}(\mathcal{W})$, such that these robot positions at time $t$ can be treated as random samples generated according to $\wp(\mathbf{x},t)$, where $\mathscr{P}(\mathcal{W})$ is the space of PDFs defined on the ROI $\mathcal{W}$. Denote the initial and target PDFs of robot positions by $\wp_0$ and $\wp_{targ}$, respectively. Then, the macroscopic objective of the VLSR system in this paper is to generate and adaptively update a trajectory of PDFs of robot positions from $\wp_0$ to $\wp_{targ}$ while preserving the robots from collisions with each other and obstacles which are incrementally observed and updated. Let $t_f$ denote the final time of the task. This macroscopic objective can be achieved by minimizing the following cost function,
\begin{equation}
J=\phi[\wp(t_f),\wp_{targ}]+\int_{t_0}^{t_f}\mathscr{L}[\wp(\mathbf{x},t),m(\mathbf{x},t)]dt
\label{eq:costFunc_cont}
\end{equation}
where the functional terms, $\phi[\wp(t_f),\wp_{targ}]$ and  $\mathscr{L}[\wp(\mathbf{x},t),m(\mathbf{x},t)]$, indicate the final term and the Lagrangian term, respectively. Because the layout of obstacles is updated according to the observations of sensors incrementally,  it is not guaranteed that the task can be completed within a given time period $[t_0, t_f]$. In other words, the final time step, $t_f$, cannot be given in advance. The task stops if and only if the robots achieve the target distribution.

\section{Background on Occupancy Mapping}
\label{sec:Distributed_Mapping}

As mentioned in Section \ref{sec:Problem_Formulation}, the approximate layout of obstacles is represented by a time-varying obstacle map function, $m(\cdot,t) \in \mathscr{M}$. There are several options for defining this map function, including traditional occupancy, the Gaussian process occupancy map, and the Hilbert occupancy map. For simplicity, the map function $m(\mathbf{x},t)$ at a certain time $t$ is defined as a binary occupancy map defined on the ROI $\mathcal{W}$, which is generated from the Hilbert occupancy map $h(\mathbf{x},t) \in [0,\, 1]$ \cite{RamosHilbertMap2016}.

A Hilbert occupancy map is a continuous probability map developed by formulating the mapping problem as a binary classification task. Let $\mathbf{x} \in \mathcal{W}$ be any point in ROI and $Y = \{0,1 \}$  be defined as a categorical random variable reflecting if the position $\mathbf{x}$ is occupied by obstacles, such that
\begin{equation}
Y = \begin{cases}
1, \quad \text{if } \mathbf{x} \in \mathcal{B}\\
0, \quad \text{otherwise}
\end{cases}
\end{equation} 
The Hilbert map $h(\mathbf{x},t)$ describes the probabilities $P(Y = 1|\mathbf{x}\in\mathcal{W})$ at the position x at the certain time $t$, as follows,
\begin{equation}
h(\mathbf{x},t) = P(Y = 1|\mathbf{x}\in\mathcal{W}) = \frac{e^{f(\mathbf{x},t)}}{1 + e^{f(\mathbf{x},t)}}
\end{equation}
where $P(Y = 1|\mathbf{x}\in\mathcal{W})$ indicates the probability that the location $\mathbf{x}\in \mathcal{W}$ is occupied by obstacles, and  $f(\mathbf{x},t)$ is a function defined on $\mathcal{W}$ at the certain time $t$. The function  $f(\mathbf{x},t)$ is learned and updated from all obtained observations up to time $t$ by minimizing a loss function of negative log-likelihood (NLL). Since the Hilbert occupancy map learning is not the key contribution, its implementation details are omitted in this paper. The interested readers are referred to \cite{ZhuGDM2019,MorelliMapping2019} for more details. Based on the updated Hilbert occupancy map $h(\mathbf{x},t)$, the obstacle map function is defined as a binary function, such that
\begin{equation}
m(\mathbf{x},t) = 
\begin{cases}
1, \quad \text{if } h(\mathbf{x},t) > 0.5 \\
0, \quad \text{otherwise}
\end{cases}
\label{eq:map_function}
\end{equation}  

\begin{figure*}[htp]
	\centering
	\subfloat[]{\includegraphics[width=2in]{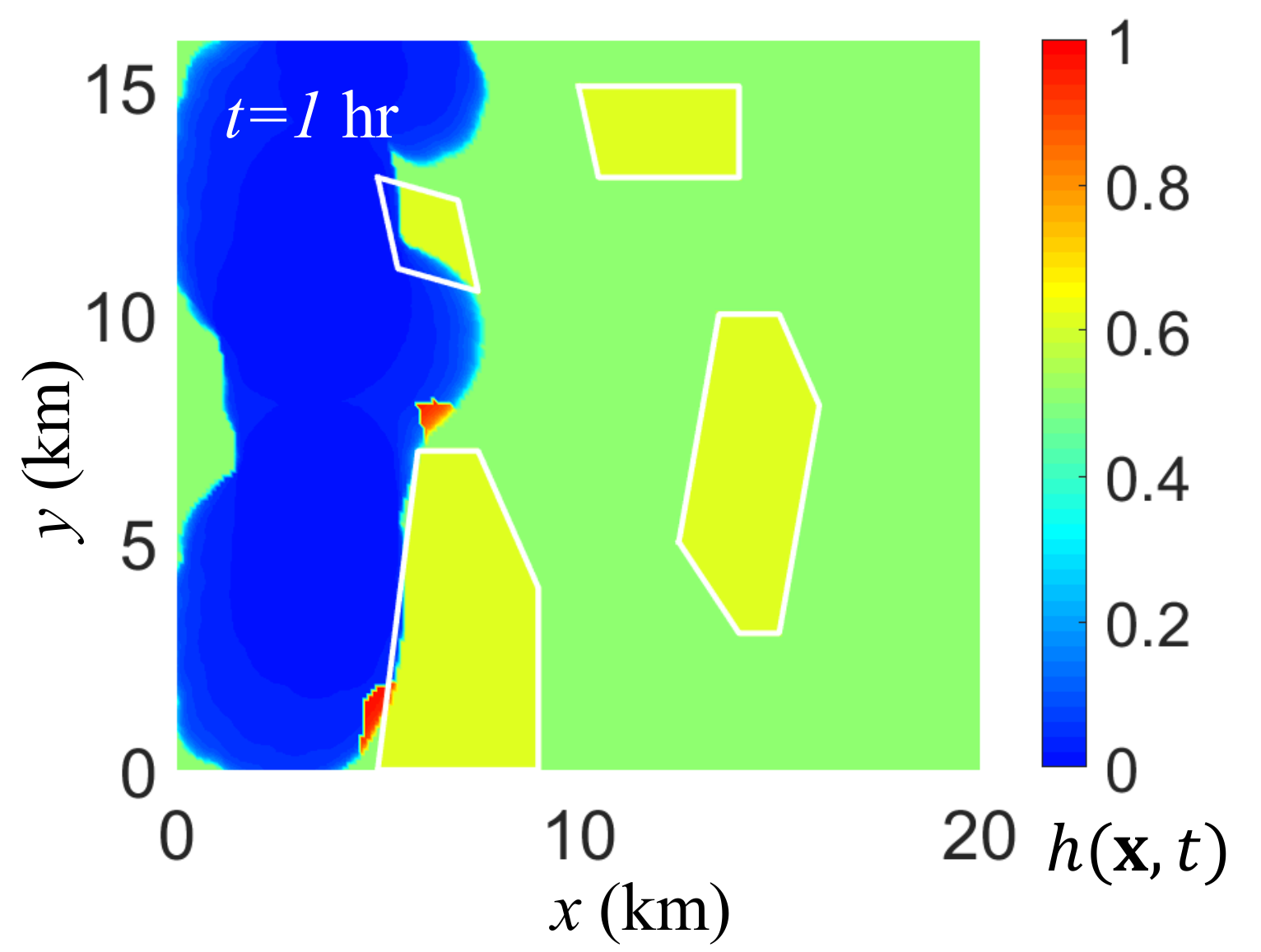}}
	\hfil
	\subfloat[]{\includegraphics[width=2in]{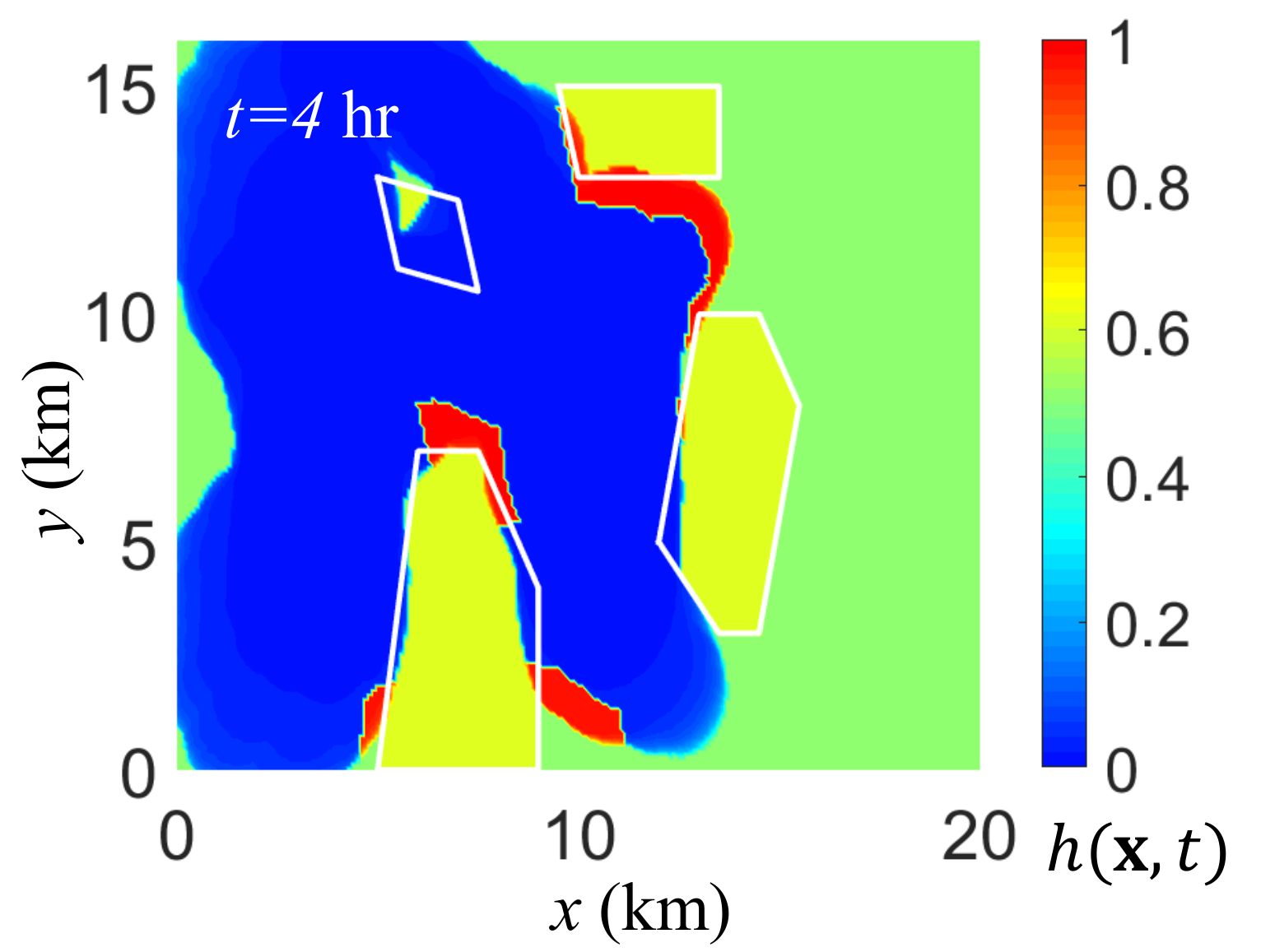}}
	\hfil
	\subfloat[]{\includegraphics[width=2in]{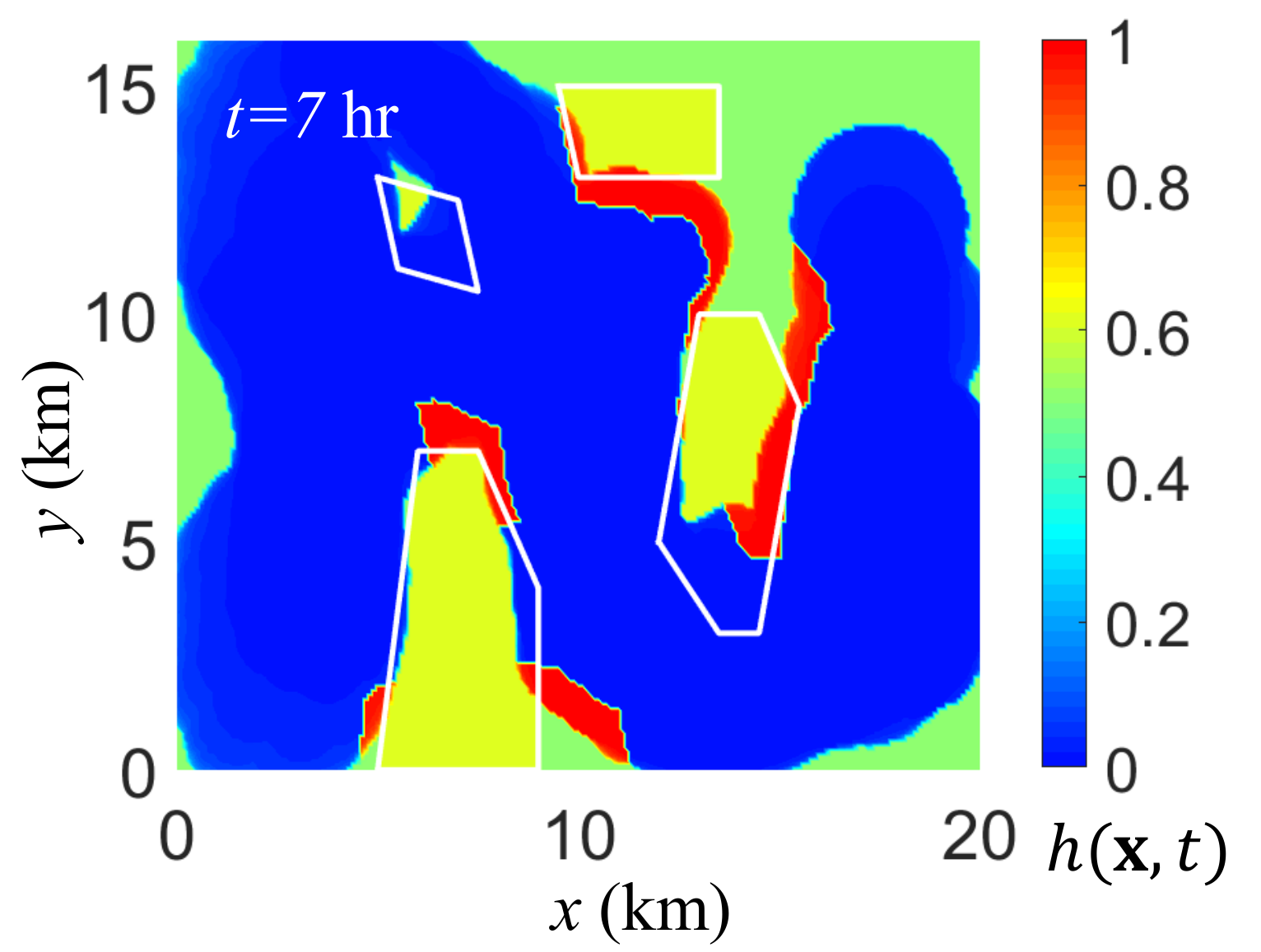}}
	\hfil
	\subfloat[]{\includegraphics[width=2in]{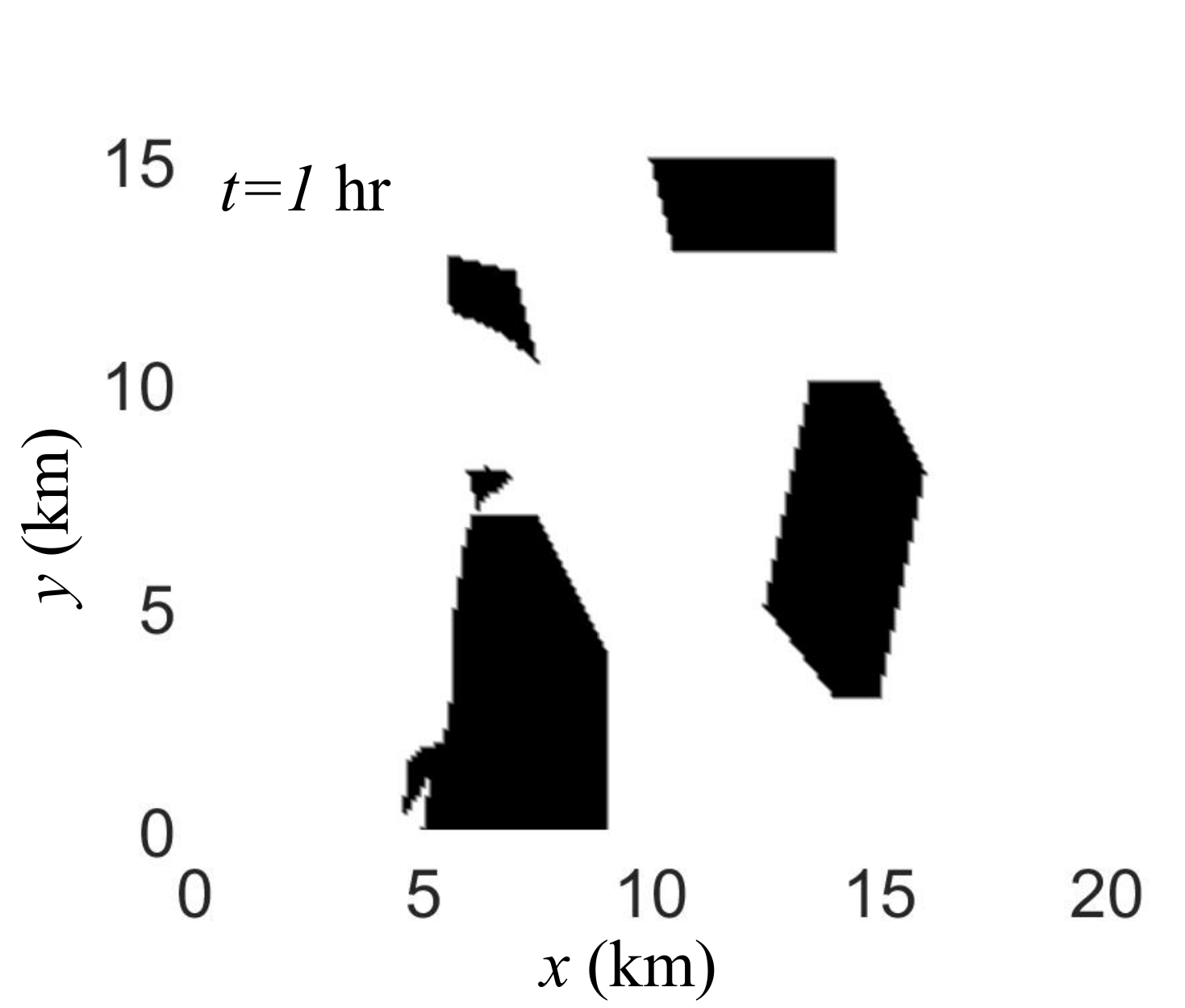}}
	\hfil
	\subfloat[]{\includegraphics[width=2in]{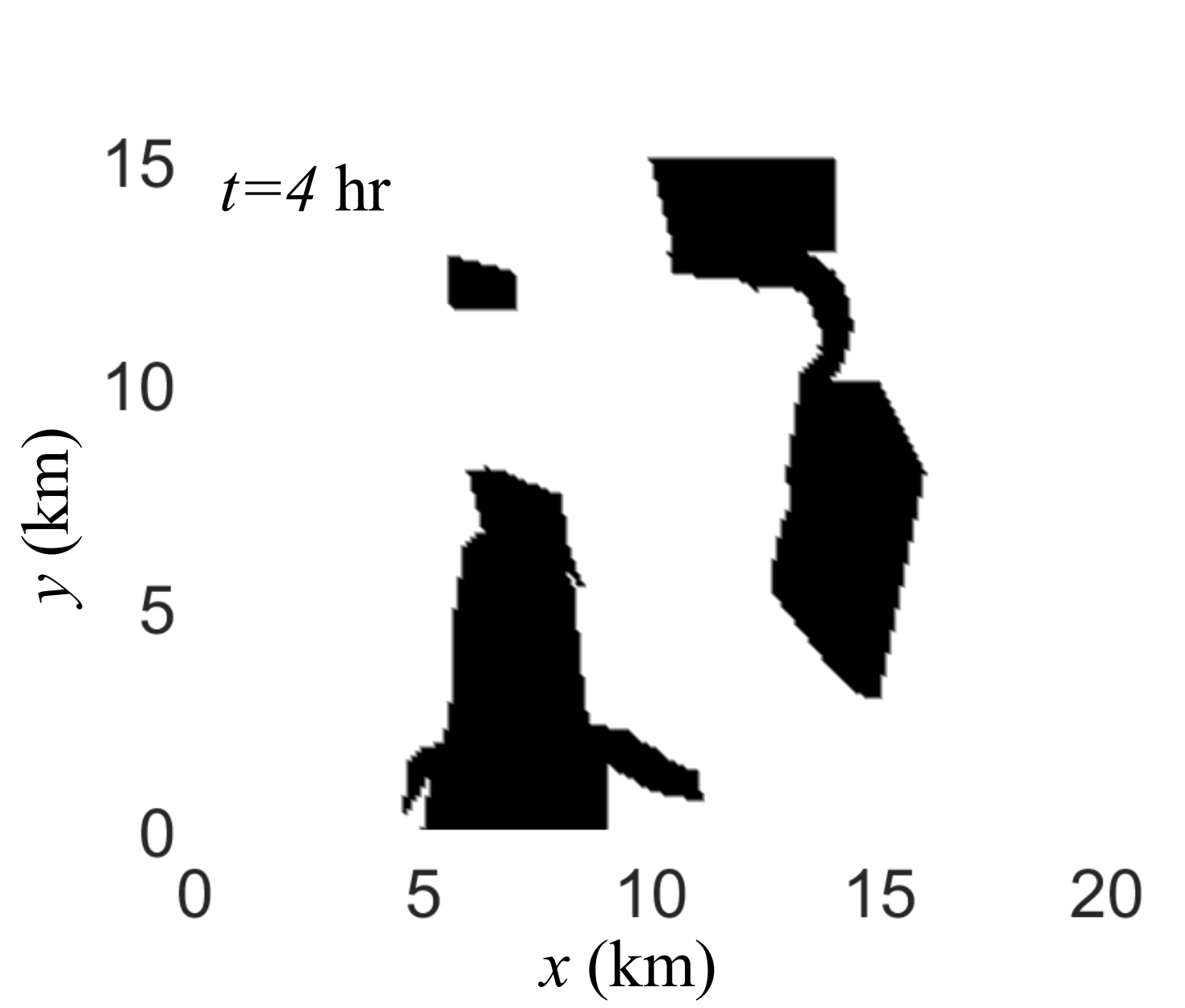}}
	\hfil
	\subfloat[]{\includegraphics[width=2in]{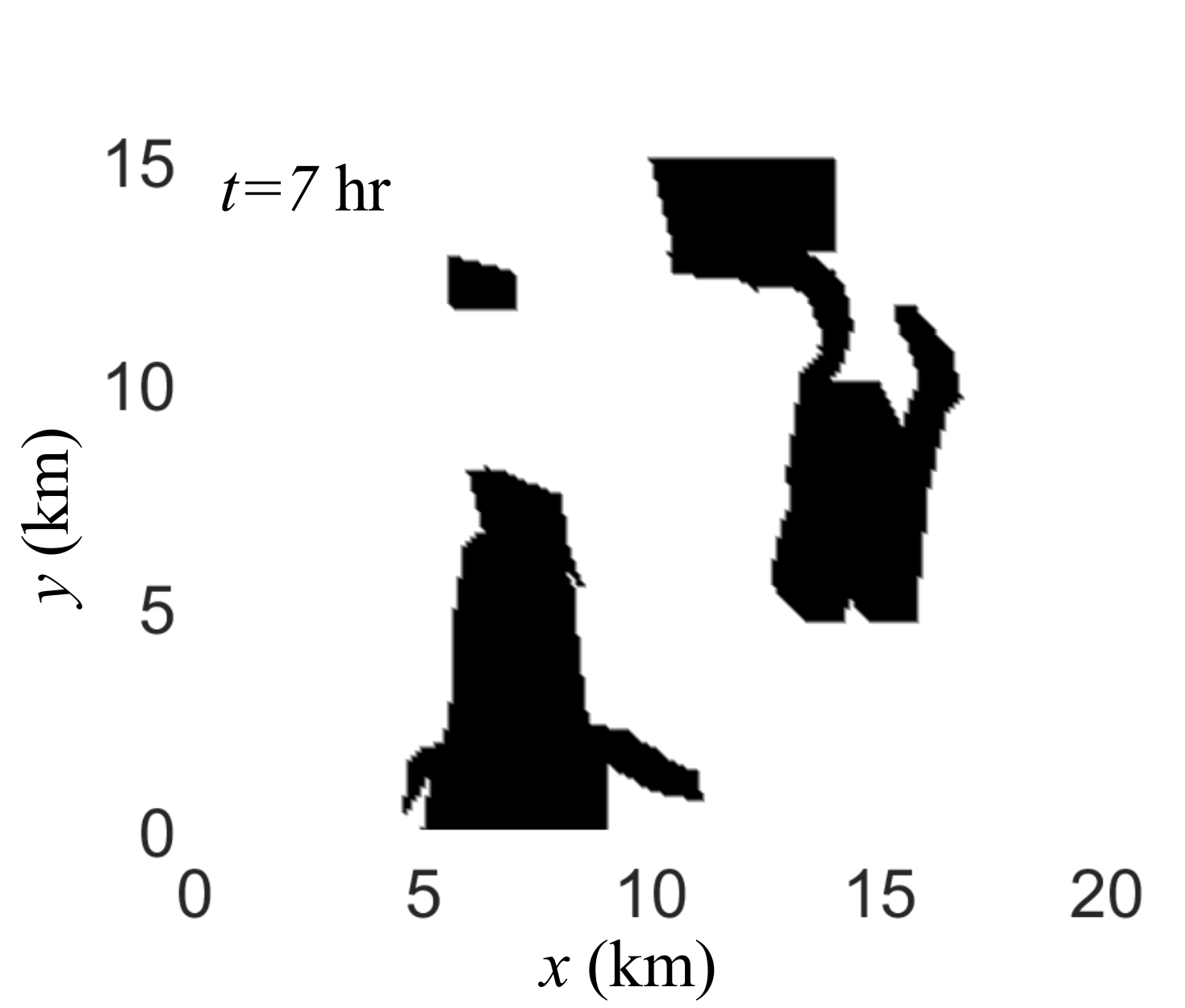}}
	\caption{The examples of the Hilbert occupancy map $h(\mathbf{x},t)$ and the corresponding obstacle map functions $m(\mathbf{x},t)$ at three different times. The sub-figures on the first row (a)-(c) represent the Hilbert occupancy map, and the sub-figures on the second row (d)-(f) represent the binary obstacle map function. The white polygons on the first row indicate the initial knowledge of the obstacles. }
	\label{fig:HM}
\end{figure*}

The example of the Hilbert function occupancy map $h(\mathbf{x},t)$ and the corresponding obstacle map functions $m(\mathbf{x},t)$ at three different times are  presented in Fig. \ref{fig:HM}. 

\section{Adaptive DOC Approach}
\label{sec:Adaptive_DOC_Approach}

In this section, the problem described in (\ref{eq:costFunc_cont}) is discretized in the temporal scale, and an adaptive DOC approach is proposed in the discrete time framework. Given a fixed time interval $0 < \triangle t \ll (t_f - t_0)$,  the task time interval $[t_0,t_f]$ can be discretized into $T_f = (t_f - t_0) / \triangle t$ equal time steps. Let $t_k = t_0 + k \triangle t$ denote the discrete-time index with $k=1,\ldots,T_f$. Then, like the DOC approach \cite{RuddIndirectDOC2013,FoderaroDOC2014,FoderaroDOC2016,FerrariDOCtutorial2016,RuddIndirectDOC2017}, the macroscopic cost function in (\ref{eq:costFunc_cont}) can be rewritten by
\begin{equation}
J=\phi(\wp_{T_f},\wp_{targ})+\sum_{k=0}^{T_f - 1} \mathscr{L}(\wp_k,m_k)
\label{eq:discrete_cost_func}
\end{equation}   
where $\wp_k = \wp(\mathbf{x},t_k)$ and $m_k = m(\mathbf{x},t_k)$ are both functions defined on $\mathcal{W}$, and $\wp_{0}$ indicates the initial robot PDF.

Because the obstacle map function is time-varying and cannot be predicted, the macroscopic objective described in (\ref{eq:discrete_cost_func}) cannot be solved by directly optimizing the trajectories of robot PDFs to minimize the objective function like the DOC approaches \cite{RuddIndirectDOC2013,FoderaroDOC2014,FoderaroDOC2016,FerrariDOCtutorial2016,RuddIndirectDOC2017}. In this paper, the optimization of the trajectory of robot PDFs is modeled as an RL-ADP problem \cite{LewisRLAPD2013}, and the objective function is reformulated by
\begin{equation}
J \triangleq \phi(\wp_{T_f},\wp_{targ} ) + \sum_{k=0}^{T_f -1} \mathscr{L} \left[\wp_k,m_k,\mathcal{C}(\wp_k,m_k)\right] 
\label{eq:RL_cost_func}
\end{equation}           
where $\mathcal{C} : \mathscr{P} \times \mathscr{M} \mapsto \mathscr{P}$ is the functional control law, such that
\begin{equation}
\wp_{k+1} \triangleq \mathcal{C}(\wp_k,m_k)
\label{eq:functional_control_law}
\end{equation}
In addition, the final time step $T_f$ is not specified in advance. The final time step $T_f$ tends to infinity if the task cannot be completed.  

It is worthy to note that because the output of the functional control law in (\ref{eq:functional_control_law}) is the robot PDF at the $(k+1)$th time step, $\wp_{k+1}$ is the functional control and is also the next functional macroscopic state in the VLSR system. In the rest of this paper, the term $\mathscr{L} \left[\wp_k,m_k,\mathcal{C}(\wp_k,m_k)\right]$ is abbreviated as $\mathscr{L} (\wp_k,m_k,\mathcal{C})$. 

The goal of the RL-ADP problem is to adaptively determine an optimal functional control law of robot PDFs to minimize the objective function defined in (\ref{eq:RL_cost_func}). Specifically, the functional control law, $\mathcal{C} (\cdot,m_k) : \mathscr{P} \mapsto \mathscr{P}$, is updated online according to the obstacle map function $m_k$ at the $k$th time step. Then, an optimal robot PDF $\wp_{k+1}$ is generated by the updated functional control law, according to (\ref{eq:functional_control_law}). Finally, the robots can navigate from the initial distribution to target distribution according to the corresponding robot PDF $\wp_{k+1}$ by utilizing the microscopic control, which will be described in Section \ref{subsec:Microscopic_Control_Law}.   

\subsection{Value Functional}
\label{subsec:Value_Function}
Unlike the traditional RL-ADP problem \cite{LewisRLAPD2013}, the functional control law in (\ref{eq:RL_cost_func}) is also dependent on the obstacle map function $m_k$ at the $k$th time step, which is updated by the incremental observations from all sensors. Assume that the obstacle map functions are static if the update of obstacles is not available. Thus, let $\mathbf{M}_k = [m_0,\ldots,m_{k-1},m_k,\ldots,m_k]$ denote a $1 \times T_f$ vector of obstacle map functions obtained at the $k$th time step. Here, $\mathbf{M}_k(\tau) = m_{\tau}$, $0 \leq \tau \leq k$, represent the observed and updated obstacle map functions up to the $k$th time step, and $\mathbf{M}_k(\tau) \equiv m_k$, $k < \tau \leq T_f-1$, represent the predicted obstacle map functions after the $k$th time step. 
Let $\mathcal{C}_{\tau}$, $\tau \leq k$, indicate the updated control law at the $\tau$th time step.
Thus, at the $k$th time step, the value functional associated to the control law $\mathcal{C}_k$ can be defined by   
\begin{align}
& \mathcal{V}_k(\wp_l,\mathbf{M},\mathcal{C}_k,l \vert \mathcal{C}_0,\ldots,\mathcal{C}_{k-1})  \nonumber\\
\triangleq & \begin{cases}
\phi(\wp_{T_f},\wp_{targ}) + \sum_{\tau=l}^{k -1} \mathscr{L}(\wp_{\tau},\mathbf{M}(\tau),\mathcal{C}_{\tau})\\
+ \sum_{\tau=k}^{T_f -1} \mathscr{L}(\wp_{\tau},\mathbf{M}(\tau),\mathcal{C}_k),
& 0 \leq l < k \\
\\
\phi(\wp_{T_f},\wp_{targ}) + \sum_{\tau=l}^{T_f -1} \mathscr{L}(\wp_{\tau},\mathbf{M}(\tau),\mathcal{C}_k), &  
k \leq l < T_f \\
\\
\phi(\wp_{T_f},\wp_{targ}), & l=T_f
\end{cases} 
\label{eq:VF_m_k}    
\end{align}
where $\mathbf{M}$ indicate the vector of obstacle map functions at any time step. 
For simplicity, the value functional is abbreviated as $\mathcal{V}_k(\wp_l,\mathbf{M})$ or $\mathcal{V}_k(\wp_l,\mathbf{M}, \mathcal{C}_k)$. 
Besides, because $\mathbf{M}_k(\tau) \equiv m_k$ for $k \leq \tau$, the term $\mathcal{V}_k(\wp_l,\mathbf{M}_k)$ is replaced by $\mathcal{V}_k(\wp_l,m_k)$ or $\mathcal{V}_k(\wp_l,m_k,\mathcal{C}_k)$ for $k \leq l < T_f$. 
 
Furthermore, a state-action functional, referred to as the Q-functional, is defined for $k \leq l < T_f$, such that
\begin{align}
\mathcal{Q}_k(\wp_l,m_k,\wp_{l+1}) &\triangleq \mathscr{L}(\wp_l,m_k,\wp_{l+1}) + \mathcal{V}_{k}(\wp_{l+1},m_k), \nonumber\\ 
&  k \leq l < T_f 
\label{eq:Q_m_k}
\end{align}
The Q-functional is a prediction of the cost-to-go from the robot PDF $\wp_l$ to $\wp_{T_f}$ at the $k$th time step, where the obstacle map function is fixed as $m_k$ and  the functional control law $\mathcal{C}_k(\cdot,m_k)$ is applied.

\subsection{Derivation of Optimal Functional Control Law}

Although the functional control law $\mathcal{C}_k$ is an operator as described before and the functional operator learning has been studied in \cite{ZhuVFA2016,ZhuMultiKernelProbDistRegression2015}, it is still challenging
to approximate this operator online and approximate $\wp_{k+1}$ according to (\ref{eq:functional_control_law}). Thus, in this paper, the critic-only Q-learning (CoQL) method is applied to obtain the optimal control where the approximation of control law $\mathcal{C}_k$ is not required \cite{LuoCoQL2016}.
Let the superscript ``*" indicate the optimal terms. According to the Bellman equation, the optimal Q-function can be expressed by
\begin{align}
\mathcal{Q}_k^* (\wp_l,m_k,\wp^*_{l+1}) &= \min_{\wp_{l+1}} \left[\mathscr{L}(\wp_l,m_k,\wp_{l+1}) + \mathcal{V}_k^* (\wp_{l+1},m_k) \right], \nonumber\\
& k\leq l < T_f
\end{align}
Thus, the output of the optimal functional control law, $\wp^{*}_{k+1}$, can be obtained by
\begin{align}
\wp^{*}_{k+1} &= \mathcal{C}_k^*(\wp_k,m_k) \nonumber\\&
 = \underset{\wp_{k+1}}{\arg\min} \left[ \mathcal{Q}_k^* (\wp_k,m_k,\wp_{k+1})\right]\nonumber\\
& =\underset{\wp_{k+1}}{\arg\min} \left[ \mathscr{L}(\wp_k,m_k,\wp_{k+1}) + \mathcal{V}_k^* (\wp_{k+1},m_k)\right]
\label{eq:optimization_optimal_control}
\end{align}
where $\mathcal{C}_k^*$ indicate the optimal control law obtained the $k$th time step. 
\subsection{Analysis of Convergence}

Given the vector of obstacle map functions at the $k$th time step $\mathbf{M}_k$, the lower bound of the optimal value functions $\mathcal{V}_{k^{\prime}}^*(\wp_l,\mathbf{M}_k)$, $0 \leq k^{\prime} \leq k$, is provided by the following theorem. 

\begin{theorem}[Lower bound of optimal control law]
	\label{theorem: optimal_control_law_lower_bound}
	At the $k$th time step, $0 < k < T_f$, given the vector of obstacle map functions $\mathbf{M}_k$, the optimal value functional $\mathcal{V}_k^*(\wp_l,\mathbf{M}_k)$ is the lower bound of all of the previous optimal value functionals $\mathcal{V}_{k^{\prime}}^*(\wp_l,\mathbf{M}_k)$ obtained at the $k^{\prime}$th time step, $0 \leq k^{\prime}\leq k$, for all robot PDFs $\wp_l$, such that
	\begin{equation}
	\mathcal{V}_k^*(\wp_l,\mathbf{M}_k) \leq \mathcal{V}_{k^{\prime}}^*(\wp_l,\mathbf{M}_k), \;0 \leq k^{\prime} \leq k \text{ and } 0 \leq l \leq T_f 
	\end{equation}
\end{theorem}
The proof of Theorem \ref{theorem: optimal_control_law_lower_bound} is provided in Appendix \ref{Appedix:lower_bound_of_optimal_control_law}.


Let $\tilde{J}(k)$, $0 \leq k <T_f$, denote the cost function with respect to the functional control law $\mathcal{C}_k$, such that
\begin{equation}
\tilde{J}(k) \triangleq \mathcal{V}_{k} (\wp_0,\mathbf{M}_{T_f-1})
\label{eq:tilde_V}
\end{equation}
%
According to the definition of the cost function in (\ref{eq:RL_cost_func}), it can be found that 
\begin{equation}
J = \tilde{J}(T_f-1) = \mathcal{V}_{T_f-1} (\wp_0,\mathbf{M}_{T_f-1})
\end{equation}

If the optimal value function $\mathcal{V}_k^*$ is available at the $k$th time step, the optimal control $\wp_{k+1}^*$ can be obtained using (\ref{eq:optimization_optimal_control}). Then, the optimal functional control law $\mathcal{C}_k^*$ is updated for $k=1,\ldots,(T_f-1)$. 
The following corollary shows that the optimal cost function $\tilde{J}^*(k)$ converges to the minimum cost function $J^*$ for $0\leq k <T_f$. 

\begin{corollary}[Convergence of cost function prediction]
	\label{corollary:Convergence_cost_function_prediction}
	The optimal cost function $\tilde{J}^*(k)$, $0 \leq k < T_f$ monotonically  converges to  $\tilde{J}^*(T_f - 1)$. For any two optimal control laws, $\mathcal{C}_k^*$ and $\mathcal{C}_{k^\prime}^*$, obtained at the $k$th and $k^{\prime}$th time steps, respectively, $0\leq k^{\prime} \leq k \leq (T_f - 1)$, such that
	\begin{equation}
	  J^* = \tilde{J}^*(T_f - 1) \leq \tilde{J}^*(k) \leq \tilde{J}^*(k^{\prime})
	\end{equation}
\end{corollary}
According to (\ref{eq:tilde_V}) and \textit{Theorem} \ref{theorem: optimal_control_law_lower_bound}, \textit{Corollary} \ref{corollary:Convergence_cost_function_prediction} can be proved straightforwardly. Therefore, by applying the optimal functional control law in (\ref{eq:optimization_optimal_control}), the cost function is minimized at each time step until the time step $(T_f - 1)$. 

\section{Background on optimal Mass Transport}
\label{sec:OMT}
Because Optimal mass transport (OMT) can deal with problems of transporting masses from an initial
distribution to a terminal one in a mass preserving manner with minimum cost, it can be applied to our proposed problem to measure the energy cost of the VLSR system from the current distribution to the next distribution. In this section, some critical concepts and properties in OMT are introduced briefly. The interested readers are referred to \cite{VillaniTopicsInOptimalTransportation2003, ChenOptimalTransportGMM2018} for more
details.

\subsection{Wasserstein Metric}
Let $\wp_p, \wp_q \in \mathcal{P}(\mathcal{W})$ denote two PDFs defined on $\mathcal{W}$, and let $\Pi(\wp_p,\wp_q)\subset\mathcal{P}(\mathcal{W}\times\mathcal{W})$
denote the set of all joint PDF $\pi\in\mathcal{P}(\mathcal{W}\times\mathcal{W})$
such that their marginals measures along the two coordinate directions
coincide with $\wp_p$ and $\wp_q$, respectively, such that
\begin{align}
&\Pi(\wp_p,\wp_q) \triangleq  \bigg\{ \pi\in\mathcal{P}(\mathcal{W}\times\mathcal{W}), \nonumber\\ 
&\int_{\mathbf{x}^{\prime}\in\mathcal{W}}\pi(\cdot,\mathbf{x}^{\prime})d\mathbf{x}^{\prime}  =\wp_p, \text{ and } \int_{\mathbf{x}\in\mathcal{X}}\pi(\mathbf{x},\cdot)d\mathbf{x}  = \wp_q\bigg\}
\end{align}
%
%
The Wasserstein
metric $W_{2}(\wp_p,\wp_q)$ is defined by \cite{VillaniTopicsInOptimalTransportation2003}
\begin{equation}
W_{2}(\wp_p,\wp_q)\triangleq\left[\underset{\pi\in\Pi(\wp_p,\wp_q)}{\inf}\int_{\mathcal{X}\times\mathcal{X}}\Vert\mathbf{x}-\mathbf{x}^{\prime}\Vert^{2}d\pi(\mathbf{x},\mathbf{x}^{\prime})\right]^{1/2}
\label{eq:W2_metric}
\end{equation}
where $\Vert \cdot \Vert$ indicate the Euclidean distance.  

It has been shown in \cite{ChenOptimalTransportGMM2018} that if both of the marginals $\wp_p \sim\mathcal{N}(\mu_{p},\Sigma_{p})$ and $\wp_q \sim\mathcal{N}(\mu_{q},\Sigma_{q})$ are
Gaussian distributions, the Wasserstein metric $W_{2}(\wp_p,\wp_q)$ can be expressed
in a closed form, such that
\begin{align}
&W_{2}(\wp_p,\wp_q)= \bigg\{ \Vert\mu_{p}-\mu_{q}\Vert^{2}  \nonumber\\
& +tr\left[\Sigma_{p}+\Sigma_{q}-2\left(\Sigma_{p}^{1/2}\Sigma_{q}\Sigma_{p}^{1/2}\right)^{1/2}\right]\bigg\} ^{1/2}
\end{align}
where $tr[\cdot]$ indicates the trace operator.  

There exists a consequent displacement interpolation $\wp_s(\epsilon)\in\mathcal{P}(\mathcal{W})$,
$\epsilon\in[0,1]$, between $\wp_p$ and $\wp_q$, where $\wp_s(0)=\wp_p$ and $\wp_s(1)=\wp_q$,
and $W_{2}(\wp_s(\epsilon),\wp_p)=\epsilon\cdot W_{2}(\wp_p,\wp_q)$ and $W_{2}(\wp_s(\epsilon),\wp_q)=(1-\epsilon)\cdot W_{2}(\wp_p,\wp_q)$ \cite{ChenOptimalTransportGMM2018}.
If $\wp_p$ and $\wp_q$ are both Gaussian distributions, then $\wp_s(\epsilon) \sim \mathcal{N}(\mu_{s}(\epsilon),\Sigma_{s}(\epsilon))$ is
also Gaussian distributed, such that
\begin{align}
\mu_{s}(\epsilon)  &= (1-\epsilon)\mu_{p}+\epsilon \mu_{q}\label{eq:mean_displacement}\\
\Sigma_{s}(\epsilon) & =  \Sigma_{p}^{-1/2}  \left[(1-\epsilon)\Sigma_{p}+\epsilon \left(\Sigma_{p}^{1/2}\Sigma_{q}\Sigma_{p}^{1/2}\right)^{1/2}\right]^{2}\Sigma_{p}^{-1/2}\label{eq:covariance_displacement}
\end{align}

\subsection{Metric on Space of Gaussian Mixture Model}

Although the Wasserstein metric $W_{2}$ can be calculated efficiently in closed
form when $\wp_p$ and $\wp_q$ are both Gaussian distributions, there is
no such closed form for general distributions, even if $\wp_p$ and $\wp_q$ are both GMMs \cite{ChenOptimalTransportGMM2018,AuricchioWassersteinMetricComputing2018}. 

Assume that $\wp_p$ and $\wp_q$ are both GMM, which can be expressed by
\begin{align}
\wp_p&=\sum_{i=1}^{N_p}\omega^{i}_{p}g_p^{i}, \quad i=1,\ldots,N_p \\
\wp_q&=\sum_{j=1}^{N_q}\omega^{j}_{q}g_q^{j}, \quad j=1,\ldots,N_q
\end{align}
where $g_p^i \sim\mathcal{N}(\mu_{p}^i,\Sigma_{p}^i)$ and $g_q^j \sim\mathcal{N}(\mu_{q}^j,\Sigma_{q}^j)$ denote Gaussian distributions, $N_p$ and $N_q$ are the numbers of Gaussian components, and $\boldsymbol{\omega}_p=[\omega^{1}_p,\ldots,\omega^{N_p}_p]$ and $\boldsymbol{\omega}_q=[\omega^{1}_q,\ldots,\omega^{N_q}_q]$ are corresponding positive probability vectors, such that $\sum_{i=1}^{N_p}\omega^{i}_p=1$ and $\sum_{j=1}^{N_q}\omega^{j}_q=1$ \cite{BishopPRML2006}. Let the space of all Gaussian mixture distributions defined on $\mathcal{W}$ be denoted by $\mathcal{G}(\mathcal{W})$.   


Recently, a new metric on $\mathcal{G}(\mathcal{W})$ was proposed in \cite{ChenOptimalTransportGMM2018} to approximate $W_{2}(\wp_p,\wp_q)$
for $\wp_p,\wp_q\in\mathcal{G}(\mathcal{W})$, such that 
%
\begin{equation}
d(\wp_p,\wp_q) \triangleq \left\{\underset{\pi\in\Pi(\boldsymbol{\omega}_{p},\boldsymbol{\omega}_{q})}{\min}\sum_{i=1}^{N_{p}}\sum_{j=1}^{N_{q}}[W_{2}(g^{i}_{p},g^{j}_{q})]^2\pi(i,j)\right\}^{1/2}
\label{eq:Approimate_WG_Metric}
\end{equation}
where $\Pi(\boldsymbol{\omega}_{p},\boldsymbol{\omega}_{q})$ denotes
the space of joint probability distributions between $\boldsymbol{\omega}_{p}$
and $\boldsymbol{\omega}_{q}$. 
%
It has been proved that $d(\cdot,\cdot)$ defines a metric on $\mathcal{G}(\mathcal{W})$
in \cite{ChenOptimalTransportGMM2018}.
In this paper, this metric is referred to as Wasserstein-GMM (WG)
metric and the GMM space associated with this metric is referred to as Wasserstein-GMM space.

Furthermore, the geodesic connecting $\wp_p$ and $\wp_q$ is given by
\begin{equation}
\wp_s(\epsilon)=\sum_{i,j}\pi^{*}(i,j)g^{ij}_{pq}(\epsilon), \quad 0\leq \epsilon \leq 1
\label{eq:distribution_interpolation}
\end{equation}
where $\pi^*(i,j)$ is the optimal joint distribution defined in (\ref{eq:Approimate_WG_Metric}), and   $g^{ij}_{pq}(\epsilon) \sim\mathcal{N}\left(\mu^{j}_{q},\Sigma^{j}_{q}\right)$ is a consequent displacement interpolation between $g^{i}_{p}$
and $g^{j}_{q}$, 
which can be specified 
according to ($\text{\ref{eq:mean_displacement}}$) and ($\text{\ref{eq:covariance_displacement}}$).

\section{Adaptive Distributed Optimal Control Based in Wasserstein-GMM Space}
\label{sec:ADOC_WG}

Considering two sequential robot PDFs, $\wp_k$ and $\wp_{k+1}$, the Wasserstein metric $W_2(\wp_k,\wp_{k+1})$ indicates the distance between these two distributions. Since the time interval $\triangle t$ between two distribution is fixed, $W_2(\wp_k,\wp_{k+1})$ is proportional to the distribution velocity at time step $k$ defined by
\begin{equation}
    \nu^{\wp}_{k} = \frac{W_2(\wp_k,\wp_{k+1})}{\triangle t}
    \label{eq:distribution_velocity_W2}
\end{equation}
Moreover, the squared of $W_2$ is proportional to the energy-cost from $\wp_k$ to $\wp_{k+1}$, which is denoted by $E_k$, such that
\begin{equation}
    E_k \propto (\nu^{\wp}_k)^2 \propto [W_2(\wp_k,\wp_{k+1})]^2
    \label{eq:Engery_cost}
\end{equation}

Furthermore, because the WG metric $d(\wp_k,\wp_{k+1})$ can be obtained via a linear optimization in  (\ref{eq:Approimate_WG_Metric}), it is reasonable to apply  $[d(\wp_k,\wp_{k+1})]^2$ as an energy-cost term in the Lagrangian term  (\ref{eq:RL_cost_func}). Compared with the divergences, such as Kullback–Leibler (KL) divergence and Cauchy-Schwarz (CS) divergence,  WG metric is more suitable for this problem because of its clear physical meaning of energy-cost and metric properties\cite{ChenOptimalTransportGMM2018}.
%

\subsection{Value functional in Wasserstein-GMM Space}

First, consider \textit{Assumption \ref{ass:wp_in_space_GMM}} presented below. 
\begin{assumption}
	\label{ass:wp_in_space_GMM}
	Assume that all of the robot PDFs belong to the GMM space, such that $\mathcal{G}(\mathcal{W})$, 
	\begin{align}
	\wp_k &= \sum_{i=1}^{N_k} \omega^i_k g^i_k, \quad k = 0,\ldots,T_f \label{eq:wp_k_in_GMM}\\
	\wp_{targ} &= \sum_{j=1}^{N_{targ}} \omega^j_{targ} g^j_{targ} \label{eq:wp_targ_in_GMM}   
	\end{align}
	where $N_k$ and $N_{targ}$ are the numbers of Gaussian components, $\boldsymbol{\omega}_{k} = [\omega^{1}_{k},\ldots,\omega^{N_{k}}_{k}]$ and $\boldsymbol{\omega}_{targ} = [\omega^{1}_{targ},\ldots,\omega^{N_{targ}}_{targ}]$ are all probability vectors, and $g^i_k$ and $g^j_{targ}$ are all Gaussian components specified by the means, $\mu^i_k$ and $\mu^j_{targ}$, and the covariance matrices, $\Sigma^i_k$ and $\Sigma^j_{targ}$, respectively.   
	Thus, the PDFs, $\wp_k$ and $\wp_{targ}$, can be specified by the tuples of parameters, $\Theta_{\wp_k} = (N_k,  \boldsymbol{g}_{k},\boldsymbol{\omega}_{k})$ and $\Theta_{\wp_{targ}}= (N_{targ}, \boldsymbol{g}_{targ}, \boldsymbol{\omega}_{targ})$, respectively, where $\boldsymbol{g}_{k} = [g^1_{k},\ldots,g^{N_{k}}_{k}]$ and $\boldsymbol{g}_{targ} = [g^1_{targ},\ldots,g^{N_{targ}}_{targ}]$ are vectors of Gaussian components. 
\end{assumption}

Then, given the robot PDF,  $\wp_k = \sum_{i=1}^{N_k} \omega^i_k g^i_k$, the functional control law  can be expressed by
\begin{align}
\wp_{k+1} &= \mathcal{C}_k(\wp_{k},m_k) \nonumber\\
&= \sum_{\imath=1}^{N_{k+1}} \omega^{\imath}_{k+1} g^{\imath}_{k+1} \nonumber\\
&= \sum_{i=1}^{N_{k}}\sum_{\imath=1}^{N_{k+1}} \pi_k(i,\imath) g^{\imath}_{k+1}
\label{eq:expression_of_control_law}
\end{align}
where $\pi_k \in \Pi(\boldsymbol{\omega}_{k},\boldsymbol{\omega}_{k+1})$
is  the  joint  probability distribution, such that 
\begin{equation}
\omega^{\imath}_{k+1} = \sum_{i=1}^{N_k} \pi_k(i,\imath), \quad \imath = 1,\ldots,N_{k+1}
\end{equation} 
Thus, given $\wp_k$ and $m_k$, the functional control law $\mathcal{C}_k$ in (\ref{eq:expression_of_control_law}) is specified by the tuple of parameters, $\Theta_{\mathcal{C}_k}=(N_{k+1}, \boldsymbol{g}_{k+1}, \pi_k )$. 

The energy-cost associated with the obstacle map function, $m_k$, and the functional control law, $\mathcal{C}_k$, can be defined by
\begin{equation}
\tilde{e}(\wp_k,m_k,\mathcal{C}_k) \triangleq \sum_{i=1}^{N_k}\sum_{\imath=1}^{N_{k+1}} [W_2(g^i_k,g^{\imath}_{k+1})]^2 \pi_k(i,\imath)
 \label{eq:tilde_d_sq_def}
\end{equation}
Like the WG metric, given $\mathcal{C}_k$ and $m_k$, a distance can be defined by
\begin{equation}
\tilde{d}(\wp_k,m_k,\mathcal{C}_k) \triangleq [\tilde{e}(\wp_k,m_k,\mathcal{C}_k)]^{1/2}
\end{equation}
According to (\ref{eq:Approimate_WG_Metric}), then, given $\wp_{k+1}$, the WG metric is the minimum of $\tilde{d}(\wp,\mathcal{C}_k)$, such that
\begin{align}
d(\wp_k,\wp_{k+1}) &=   \underset{\pi_k}{\min} \bigg\{ \sum_{i=1}^{N_k}\sum_{\imath=1}^{N_{k+1}} [W_2(g^i_k,g^{\imath}_{k+1})]^2 \pi_k(i,\imath) \bigg\}^{1/2}\nonumber\\
& = \underset{\pi_k}{\min} [\tilde{d}(\wp_k,m_k,\mathcal{C}_k)]
\label{eq:def_WG_Metric}
\end{align}

Furthermore, considering the constraint of the obstacles to the robot PDF, the Lagrangian term is defined by
\begin{equation}
	\mathscr{L}(\wp_k,m_k,\mathcal{C}_k) = [\tilde{d}(\wp_k,m_k,\mathcal{C}_k)]^2 +  \langle   \mathcal{C}_k(\wp_{k},m_k), m_k \rangle_{\mathcal{W}}
	\label{eq:Lagrangian_GW}
\end{equation}
where $\langle \cdot,\cdot \rangle_{\mathcal{W}}$
indicates the inner product on the ROI $\mathcal{W}$. Here, the second term reflects the probability that the sensors at the $(k+1)$th time step are deployed inside the approximate obstacles $\hat{\mathcal{B}}(t_k)$.

Finally, the final term in (\ref{eq:RL_cost_func}) is defined by the WG metric, such that
\begin{equation}
\phi(\wp_{T_f},\wp_{targ} ) \triangleq [d(\wp_{T_f},\wp_{targ} )]^2
\end{equation}
Therefore, the cost function in (\ref{eq:RL_cost_func}) can be rewritten by
\begin{align}
\label{eq:RL_cost_func_WG} 
J & \triangleq 
[d(\wp_{T_f},\wp_{targ} )]^2 \nonumber\\
& + \sum_{k=0}^{T_f -1} \left\{[\tilde{d}(\wp_k,m_k,\mathcal{C}_k)]^2 +  \langle   \wp_{k+1}, m_k \rangle_{\mathcal{W}} \right\} 
\end{align} 
According to (\ref{eq:VF_m_k}), thus, the corresponding value functional $\mathcal{V}_k(\wp_{k},m_k,\mathcal{C}_k)$ can be expressed by
\begin{align}
    \mathcal{V}_k(\wp_{k},m_k,\mathcal{C}_k) &= [d(\wp_{T_f},\wp_{targ} )]^2 
    + \sum_{\tau=k}^{T_f -1} \bigg\{[\tilde{d}(\wp_{\tau},m_k,\mathcal{C}_k)]^2 \nonumber\\
    &+  \langle   \mathcal{C}_k(\wp_{\tau},m_k), m_k \rangle_{\mathcal{W}} \bigg\} \nonumber\\
    &= \mathscr{L}(\wp_k,m_k,\mathcal{C}_k) + \mathcal{V}_k(\wp_{k+1},m_k,\mathcal{C}_k)
    \label{eq:VF_WG}
\end{align}

\subsection{Approximation of Optimal Value functional in Wasserstein-GMM Space}

According to (\ref{eq:VF_m_k}), the value functional $\mathcal{V}_k(\wp_{k+1},m_k,\mathcal{C}_k)$ depends on the trajectory of robot PDFs, $\{ \wp_{\tau}\}_{\tau = k+1}^{T_f}$, which are not available at the $k$th time step. Therefore, the optimal value functional  $\mathcal{V}_k^*(\wp_{k+1},m_k)$ in (\ref{eq:optimization_optimal_control}) is not available.  An approximation of the optimal value functional is required. Similar to many other RL-ADP approaches \cite{LewisRLAPD2013}, an upper bound of the optimal value functional is applied as the approximation. 

 
Consider a special control law, $\tilde{\mathcal{C}}_k(\cdot,m_k) : \mathscr{P} \mapsto \mathscr{P}$. For any time step $\tau$, $k+1 \leq \tau \leq T_f-1$,  the next step robot PDF is obtained by $\wp_{\tau+1} = \tilde{C}_k(\wp_{\tau},m_k)$, such that
\begin{align}
    N_{\tau+1} &= N_{\tau}, \quad k+1 \leq \tau \leq T_f-1 \label{eq:number_GC}\\
    \pi_{\tau}(i,\imath) &= \begin{cases}
    \omega^i_{\tau}, & \text{if } i = \imath \\
    0, & \text{otherwise}
    \end{cases}, \quad i,\imath = 1,\ldots, N_{\tau}
    \label{eq:weight_GC}
\end{align}
It means that from the ($k+1$)th time step to the final time step, the number and the corresponding weights of the Gaussian components of  $\wp_{\tau}$ are fixed. 

By recursively applying $\tilde{\mathcal{C}}_k(\cdot,m_k)$, a trajectory of robot PDFs can be generated from $\wp_{k+1}$, such that
\begin{equation}
    \wp_{\tau} = \sum_{\imath=1}^{N_{k+1}}\omega^{\imath}_{k+1} g^{\imath}_{\tau}, \quad \tau = k+1,\ldots,T_f
\end{equation}
Thus, there are $N_{k+1}$ trajectories of Gaussian components from $g^{\imath}_{k+1}$ to $g^{\imath}_{T_f}$, $\imath = 1,\ldots,N_{k+1}$. Next, consider the $N_{targ}$ Gaussian components of the target PDF, $\{g^j_{targ}\}_{j=1}^{N_{targ}}$. There are $N_{k+1} \times N_{targ}$ trajectories of Gaussian components denoted by the trajectory set $\tilde{\mathscr{Tr}}_k = \big\{(g^{\imath}_{k+1},\ldots,g^{\imath}_{T_f}, g^j_{targ})\big\}_{\imath,j}^{N_{k+1},N_{targ}}$, where each element of the set $\tilde{\mathscr{Tr}}_k$ represents a trajectory of Gaussian components denoted by $\tilde{\mathscr{Tr}}^{\imath,j}_k =  (g^{\imath}_{k+1},\ldots,g^{\imath}_{T_f}, g^j_{targ})$. Let $\tilde{\mathcal{L}}^{\imath,j}_k$ denote the minimum cost function of the trajectory of Gaussian components, $\tilde{\mathscr{Tr}}^{\imath,j}_k$, at the $k$th time step, defined by
\begin{align}
    \tilde{\mathcal{L}}^{\imath,j}_k 
    &= \begin{cases}
     \underset{\tilde{\mathscr{Tr}}^{\imath,j}_k}{\min}\big\{[W_2(g^{\imath}_{T_f},g^{j}_{targ})]^2 \\
     + \sum_{\tau = k+1}^{T_f - 1} [W_2(g^{\imath}_{\tau},g^{\imath}_{\tau+1})]^2 & \text{if } k+1 < T_f\\
     + \sum_{\tau = k+1}^{T_f - 1}\langle g^{\imath}_{\tau+1},m_k \rangle \big\},   \\
    \\
    [W_2(g^{\imath}_{T_f},g^{j}_{targ})]^2, & \text{if } k+1 = T_f
    \end{cases}
    \label{eq:L_k}
\end{align}
%

By applying  $\tilde{\mathcal{C}}_k(\cdot,m_k)$, an upper bound of the optimal value functional  $\mathcal{V}_k^*(\wp_{k+1},m_k)$ is provided in the following theorem. 
\begin{theorem}[Upper bound of optimal value functional]
    \label{theorem: Upper_bound_of_optimal_value_functional}
Given the robot PDF $\wp_{k+1}$ and the target robot PDF $\wp_{targ}$ defined in (\ref{eq:wp_k_in_GMM}) and (\ref{eq:wp_targ_in_GMM}), respectively, there exists an upper bound of the optimal value functional $\mathcal{V}_{k}^*(\wp_{k+1},m_k)$, which is denoted by $\tilde{\mathcal{V}}_{k}(\wp_{k+1},m_k)$, such that
\begin{equation}
\mathcal{V}_{k}^*(\wp_{k+1},m_k) \leq \tilde{\mathcal{V}}_{k}(\wp_{k+1},m_k) 
\label{eq:upper_bound_optimal_VF}
\end{equation}
where 
\begin{equation}
    \tilde{\mathcal{V}}_{k}(\wp_{k+1},m_k) \triangleq \sum_{\imath=1}^{N_{k+1}} \sum_{j=1}^{N_{targ}}\tilde{\mathcal{L}}^{\imath,j}_k\tilde{\pi}_{k}(\imath,j)
\label{eq:def_upper_bound_VF}
\end{equation}
and 
$\tilde{\pi}_{k}(\imath,j) \in \Pi(\boldsymbol{\omega}_{k+1},\boldsymbol{\omega}_{targ})$ is the joint probability distribution.
\end{theorem}
The proof of \textit{Theorem} \ref{theorem: Upper_bound_of_optimal_value_functional} is provided in Appendix \ref{Appedix: Upper_bound_of_optimal_value_functional}. 

In this paper, the upper bound of the optimal value functional in (\ref{eq:def_upper_bound_VF}) is applied to approximate the optimal value functional $\mathcal{V}_k^*(\wp_{k+1},m_k)$, such that
\begin{equation}
\hat{\mathcal{V}}_k^*(\wp_{k+1},m_k) 
\triangleq \sum_{\imath=1}^{N_{k+1}} \sum_{j=1}^{N_{targ}} \tilde{\mathcal{L}}^{\imath,j}_k \tilde{\pi}_{k}(\imath,j)
\label{eq:appr_optimal_VF}
\end{equation}

\subsection{Optimal Functional Control Law in Wasserstein-GMM Space}

According to (\ref{eq:VF_WG}) and (\ref{eq:appr_optimal_VF}), the optimal control law can be approximated by solving the following optimization problem,
\begin{align}
\mathcal{C}_k^* 
&\approx \underset{\mathcal{C}_k}{\arg\min} \left[\mathscr{L}(\wp_k,m_k,\mathcal{C}_k) + \hat{\mathcal{V}}_k^*(\wp_{k+1},m_k) \right] \nonumber\\
&= \underset{\mathcal{C}_k}{\arg\min} \bigg\{ [\tilde{d}(\wp_k,m_k,\mathcal{C}_k)]^2 +  \langle   \mathcal{C}_k(\wp_{k},m_k), m_k \rangle_{\mathcal{W}} \nonumber\\
&+ \sum_{\imath=1}^{N_{k+1}} \sum_{j=1}^{N_{targ}} \tilde{\mathcal{L}}^{\imath,j}_k \tilde{\pi}_{k}(\imath,j) \bigg\}
\label{eq:optimization_C_k}
\end{align}
Because the control law is specified by the tuple of parameter, $\Theta_{\mathcal{C}_k}=(N_{k+1}, \boldsymbol{g}_{k+1}, \pi_k )$ in (\ref{eq:expression_of_control_law}), and the optimal value functional is approximated by $\{ \tilde{\mathcal{L}}^{\imath,j}_k\}_{\imath,j}^{N_{k+1},N_{targ}}$ and $\tilde{\pi}_k$ in (\ref{eq:appr_optimal_VF}), the optimization problem in (\ref{eq:optimization_C_k}) can be expressed as,
\begin{align}
\hat{\Theta}^*_{\mathcal{C}_k} &= \underset{\Theta_{\mathcal{C}_k}}{\arg\min} \bigg\{  \sum_{i=1}^{N_{k}}\sum_{\imath=1}^{N_{k+1}}[W_2(g^i_k,g^{\imath}_{k+1})]^2\pi_k(i,\imath) \nonumber\\
&+ \sum_{i=1}^{N_k}\sum_{\imath = 1}^{N_{k+1}} \langle g^{\imath}_{k+1},m_k \rangle_{\mathcal{W}} \pi_k(i,\imath) \nonumber\\
&+ \sum_{\imath=1}^{N_{k+1}} \sum_{j=1}^{N_{targ}} \tilde{\mathcal{L}}^{\imath,j}_k \tilde{\pi}_{k}(\imath,j) \bigg\}
\end{align}
where $\hat{\Theta}^*_{\mathcal{C}_k}$ is the approximation of $\Theta^*_{\mathcal{C}_k}$.

There are $N_k \times N_{k+1}$ trajectories of Gaussian components denoted by the trajectory set $\mathscr{T}_k = \{(g^i_k,g^{\imath}_{k+1})\}_{i,\imath}^{N_k,N_{k+1}}$. Similarly, let $\mathcal{L}^{i,\imath}_k$ denote the cost function of the 
trajectory of Gaussian components, 
$\mathscr{T}^{i,\imath}_k = (g^{i}_k,g^{\imath}_{k+1})$,
with respect to the map function $m_k$, which is defined by
\begin{equation}
\mathcal{L}^{i,\imath}_k = [W_2(g^i_k,g^{\imath}_{k+1})]^2 + \langle g^{\imath}_{k+1}, m_k \rangle_{\mathcal{W}}
\label{eq:tilde_L_k}
\end{equation}
 
Finally, considering the following constraints of the joint probabilities, 
\begin{align}
\omega^i_k &= \sum_{\imath=1}^{N_{k+1}} \pi_k(i,\imath) \label{eq:constraint_1}\\
\omega^j_{targ} &= \sum_{\imath=1}^{N_{k+1}} \tilde{\pi}_k(\imath,j) \label{eq:constraint_2}\\
\omega^{\imath}_{k+1} &= \sum_{i=1}^{N_k} \pi_k(i,\imath) = \sum_{j=1}^{N_{targ}} \tilde{\pi}_k(\imath,j) \label{eq:constraint_3}
\end{align} 
the optimal control law can be approximated by solving the following optimization problem,

\begin{align}
\hat{\Theta}^*_{\mathcal{C}_k} &= \underset{\Theta_{\mathcal{C}_k}}{\arg\min} \sum_{\imath=1}^{N_{k+1}} \bigg [ \sum_{i=1}^{N_{k}} \mathcal{L}^{i,\imath}_k
\pi_k(i,\imath) 
+  \sum_{j=1}^{N_{targ}} \tilde{\mathcal{L}}^{\imath,j}_k \tilde{\pi}_k(\imath,j) \bigg ]\nonumber\\
\text{s.t. } & \quad (\ref{eq:constraint_1}) - (\ref{eq:constraint_3}) 
\label{eq:Theta_k_next_2}
\end{align}

After obtaining $\hat{\Theta}^*_{\mathcal{C}_k} = (\hat{N}^*_{k+1}, \hat{\boldsymbol{g}}^*_{k+1}, \hat{\pi}^*_k )$, the optimal robot PDF $\wp^*_{k+1}$ can be approximated by
\begin{equation}
\hat{\wp}^*_{k+1} = \sum_{i=1}^{N_k}\sum_{\imath = 1}^{\hat{N}^*_{k+1}}
\hat{\pi}_k^*(i,\imath) (\hat{g}^{\imath}_{k+1})^* = \sum_{\imath = 1}^{\hat{N}^*_{k+1}} (\hat{\omega}^{\imath}_{k+1})^* (\hat{g}^{\imath}_{k+1})^*
\end{equation}
where $(\hat{\omega}^{\imath}_{k+1})^* = \sum_{i=1}^{N_k} \hat{\pi}_k^*(i,\imath)$ and $\hat{\boldsymbol{g}}^*_{k+1} = [(\hat{g}^1_{k+1})^*, \ldots, (\hat{g}^{\hat{N}_{k+1}^*}_{k+1})^*]$.

Given $\hat{\wp}^*_{k+1}$ and the approximated optimal control law, $\hat{\mathcal{C}}^*_k$, which is specified by $\hat{\Theta}^*_{\mathcal{C}_k}$,  the corresponding Lagrangian term defined in (\ref{eq:Lagrangian_GW}) can be expressed by
\begin{align}
\mathscr{L}(\wp_k,m_k,\hat{\mathcal{C}}^*_k) &= [\tilde{d}(\wp_k,m_k,\hat{\mathcal{C}}^*_k)]^2 +  \langle   \hat{\mathcal{C}}^*_k(\wp_{k},m_k), m_k \rangle_{\mathcal{W}} \nonumber\\
&= \sum_{i=1}^{N_k}\sum_{\imath=1}^{\hat{N}^*_{k+1}} [W_2\big(g^i_k,(\hat{g}^{\imath}_{k+1})^*\big)]^2 \hat{\pi}^*_k(i,\imath) \nonumber\\
&+ \langle   \hat{\wp}^*_{k+1}, m_k \rangle_{\mathcal{W}} \nonumber\\
&= [d(\wp_k,\hat{\wp}^*_{k+1})]^2 + \langle   \hat{\wp}^*_{k+1}, m_k \rangle_{\mathcal{W}}
\end{align}
Thus, the Lagrangian term under $\hat{\mathcal{C}}^*_k$ is a function of the WG metric between $\wp_k$ and $\hat{\wp}^*_{k+1}$. Let $\wp_{k+1} = \hat{\wp}^*_{k+1}$ and $v^{\wp}_k$ denote the velocity of robot PDFs at the $k$th time step, defined by
\begin{equation}
v^{\wp}_k \triangleq \frac{d(\wp_k,\wp_{k+1})}{\Delta t}
\label{eq:distribution_velocity_WG}
\end{equation}  
which is an approximation of distribution velocity defined in (\ref{eq:distribution_velocity_W2}). Then, the Lagrangian term, $\mathscr{L}(\wp_k,m_k,\hat{\mathcal{C}}^*_k)$,  reflects the energy-cost $E_k$ in (\ref{eq:Engery_cost}). 

\section{Implementation of Adaptive Distributed Optimal Control}
\label{sec:Implementation}
To obtain the approximated optimal functional control law online,  an numerical approximation approach is presented in this section.

\subsection{Approximation of Optimal Control Law in Sub-space of GMM}
\label{subsec:Approximation_in_SubSpace}
According to  (\ref{eq:Theta_k_next_2}), there are three problems result in the very high computational complexity, including 
\begin{enumerate}
    \item unknown number of Gaussian components $N_{k+1}$
    \item nonlinear calculations of $W_2(\cdot,\cdot)$ in (\ref{eq:L_k}) and (\ref{eq:tilde_L_k}) 
    \item nonlinear programming (NLP) in (\ref{eq:L_k})
\end{enumerate}
To simplify these problems, the following assumption is made.
\begin{assumption}
\label{ass:collocation_GC} 
Assume that at the $k$th time step, $0\leq k < T_f$, the robot PDFs, $\wp_{\tau} \in \mathcal{G}(\mathcal{W})$, $\tau = k+1,\ldots,T_f$, can  be expressed by
\begin{equation}
    \wp_{\tau} = \sum_{\imath=1}^{N_c} \omega^{\imath}_{\tau} g^{\imath}_c
    \label{eq:wp_in_subspace}
\end{equation}
where these parameters $N_c$ and $g^{\imath}_c$, $\imath=1,\ldots,N_c$, are all fixed and known a priori, which specify a set of collocation Gaussian components denoted by $\boldsymbol{G} = \{g^1_{c},
\ldots,g^{N_{c}}_{c}\}$. Here, the subscript ``c" indicates the collocation Gaussain components. 
\end{assumption}

Specifically, these collocation Gaussian components, $g^{\imath}_c$ $\imath = 1,\ldots,N_c$, are all specified by the corresponding means $\mu^{\imath}_c$ and covariance matrices $\Sigma^{\imath}_c$. There are several methods to set these parameters, including random sampling and uniform deploying. In this paper, a common covariance matrix is set for all collocation Gaussian components, such that $\Sigma^{\imath}_c = \Sigma_c$, and the means of the collocation Gaussian components are uniformly deployed on the ROI with the same spatial interval.   
 
\textit{Assumption} \ref{ass:collocation_GC} is an extension of \textit{Assumption} \ref{ass:wp_in_space_GMM} by assuming that the Gaussian components,  
$g^{\imath}_{\tau}$, $k+1 \leq \tau \leq T_f$,
in  (\ref{eq:wp_k_in_GMM})  all belong to the set of collocation Gaussian components, such that  $g^\imath_{\tau} \in \boldsymbol{G}$. According to (\ref{eq:wp_in_subspace}), thus,  $\wp_{\tau}$ belongs to a sub-space of the GMM, $\tilde{\mathcal{G}}(\mathcal{W},\boldsymbol{G}) \subset \mathcal{G}(\mathcal{W})$, which is defined by
\begin{align}
    \tilde{\mathcal{G}}(\mathcal{W},\boldsymbol{G}) & \triangleq \bigg\{\wp \bigg\vert \wp = \sum_{\imath=1}^{N_c} \omega_{\imath} g^{\imath}_c, \sum_{\imath}^{N_c} \omega_{\imath} = 1, \nonumber\\
    & 0 \leq \omega_{\imath} \leq 1, \imath = 1,\ldots,N_c \bigg\}
\end{align}

With \textit{Assumption} \ref{ass:collocation_GC}, first, the number of Gaussian component of $\wp_{k+1}$ is known, such that $N_{k+1} = N_c$. Second, the metric $W_2(\cdot,\cdot)$ in (\ref{eq:L_k}) and (\ref{eq:tilde_L_k}) can be calculated in advance. 
Finally, due to $g^{\imath}_{\tau} \in \boldsymbol{G}$, $\tau = k+1,\ldots,T_f$, the NLP in (\ref{eq:L_k}) can be solved by using the shortest-path-planing algorithms \cite{ThorupShortestPathPlanning2004}. 

Let $\boldsymbol{V} = \boldsymbol{G} \cup \{ g^j_{targ}\}_{j=1}^{N_{targ}}$ denote a set of Gaussian components, which are treated as nodes in a graph. Let $\mathcal{E}_k$ denote a set of edges between nodes with respect to the obstacle map function $m_k$, such that 
\begin{equation}
    \mathcal{E}_k = \big \{ e_{i,j} \big \vert e_{i,j} = [W_2(g_i,g_j)]^2 + \langle g_j, m_k \rangle_{\mathcal{W}}, \text{ and } g_i,g_j \in \boldsymbol{V} \big\}
\end{equation}
Then, a directed graph $\mathcal{DG}_k = (\boldsymbol{V}, \mathcal{E}_k)$ is defined. The cost of Gaussian component trajectory, $\tilde{\mathcal{L}}^{\imath,j}_k$ in (\ref{eq:L_k}) can be explained as the shortest path from the node $g^{\imath}_{k+1} \in \boldsymbol{V} $ to the node $g_j^{targ}\in \boldsymbol{V}$ in the directed graph $\mathcal{DG}_k$.

Therefore, given the set  of  collocation Gaussian components $\boldsymbol{G}$, under \textit{Assumption} \ref{ass:collocation_GC} the  functional control law is only specified by $\pi_k$. The optimal functional control law, thus, can be approximated by solving the following optimization problem,
\begin{align}
\hat{\pi}_k^* &= \underset{\pi_k}{\arg\min} \sum_{\imath=1}^{N_{c}} \bigg[ \sum_{i=1}^{N_k} \mathcal{L}^{i,\imath}_k
\pi_k(i,\imath) 
+  \sum_{j=1}^{N_{targ}} \tilde{\mathcal{L}}^{\imath,j}_k \tilde{\pi}_k(\imath,j) \bigg]\nonumber\\
\text{s.t. } & \quad (\ref{eq:constraint_1}) - (\ref{eq:constraint_3}) 
\label{eq:Theta_k_next_3}
\end{align}
and the optimal robot PDF $\wp^*_{k+1}$ can be approximated by
\begin{equation}
\hat{\wp}^*_{k+1} = \sum_{i=1}^{N_k}\sum_{\imath =1}^{N_c} \hat{\pi}^*_k(i,\imath) g^{\imath}_c = \sum_{\imath=1}^{N_c}(\hat{\omega}_{\imath}^{k+1})^* g^{\imath}_c
\label{eq:hat_pi_k_next}
\end{equation}

In addition, given the costs of trajectories of Gaussian components, $\{\mathcal{L}^{i,\imath}_k \}_{i=1,\imath=1}^{N_k,N_c}$ and $\{\tilde{\mathcal{L}}^{\imath,j}_k\}_{\imath=1,j=1}^{N_c,N_{targ}}$, the approximated optimal value functional only depends on $\pi_k$ and $\tilde{\pi}_k$ with the constraints in (\ref{eq:constraint_1}), (\ref{eq:constraint_2}) and (\ref{eq:constraint_3}). Therefore, the optimal control law $\mathcal{C}_k^*$ can be approximated by using linear programming (LP) algorithms.

According to (\ref{eq:hat_pi_k_next}), the approximated optimal robot PDF, $\hat{\wp}_{k+1} \in \tilde{\mathcal{G}}(\mathcal{W},\boldsymbol{G})$, is specified by $N_c$ weight coefficients, $(\hat{\omega}^{\imath}_{k+1})^*$, $\imath = 1,\ldots,N_c$, where many weight coefficients are zeros or very small numbers. To reduce the computational complexity,  these Gaussian components with the weight coefficients that are smaller than a given threshold, $\omega_{th}$, are removed, and the remaining weight coefficients are normalized to generate the robot PDF at the $(k+1)$th time step, $\wp_{k+1} = \sum_{i=1}^{N_{k+1}} \omega^i_{k+1} g^{i}_{k+1} $, where $N_{k+1} \ll N_c$.

\subsection{Simplified Approximation of Optimal Control Law}
 Although the optimal $\wp_{k+1}$ can be approximated in (\ref{eq:hat_pi_k_next}) by using LP algorithms, this optimization problem cannot be implemented online with acceptable performance. Because the computational complexities of $\{\mathcal{L}^{i,\imath}_k \}_{i=1,\imath=1}^{N_k,N_c}$ and $\{\tilde{\mathcal{L}}^{\imath,j}_k\}_{\imath=1,j=1}^{N_c,N_{targ}}$ increase dramatically with the increase of the number of collocation Gaussian components.
The following assumption is considered to reduce the computational complexity of the approximation of the optimal control law. 
\begin{assumption}
	\label{ass:number_next_GC}
	  Given the robot PDF, $\wp_l = \sum_{i=1}^{N_l} \omega^i_l g^i_l$, and $\wp_{l+1}=\sum_{\imath=1}^{N_c} \omega^{\imath}_{l+1} g^{\imath}_{c} \in \tilde{\mathcal{G}}(\mathcal{W},\boldsymbol{G})$, the control law, $\wp_{l+1} = \mathcal{C}_k(\wp_l)$, $l = k, \ldots, T_f-1$, can be specified by the joint probability $\pi_l$, which is an $N_l \times N_c$ matrix. Assume that the transportation distance of Gaussian compoenents under the control law at any time step is less than a given distance threshold, $d_{th}$, such that  
	\begin{align}
	\pi_l(i,\imath) = 0  \text{ if } W_2(g^i_l, g^{\imath}_{l+1}) > d_{th}, \nonumber\\
	i = 1,\ldots,N_l, \text{ and } \imath = 1,\ldots,N_c
	\end{align} 
\end{assumption}

Let $\mathcal{I} = \{1,\ldots,N_c\}$ denote the index set of the collocation Gaussian components. For each Gaussian component of $\wp_l = \sum_{i=1}^{N_l} \omega^i_l g^i_l$, $l = k,\ldots,T_f-1$, these exists a subset of the index set $\mathcal{I}$, denoted by $\mathcal{I}^i_l \subset \mathcal{I}$, such that
\begin{equation}
\mathcal{I}^i_l = \{ \imath \vert \imath \in \mathcal{I}, \, g^{\imath}_c \in \boldsymbol{G}, \, \text{and } W_2(g^i_l, g^{\imath}_c) \leq d_{th}\}
\end{equation}  
The union of these subsets are denoted by $\mathcal{I}_l = \cup_{i = 1}^{N_l} \mathcal{I}^i_l$.
Also,  there exists a subset $\mathcal{I}^C_l \in \mathcal{I}$, such that
\begin{equation}
\mathcal{I}^C_l = \{\imath \vert \imath \in \mathcal{I} \text{ and } \imath \notin \mathcal{I}_l\}
\end{equation}
which is the complement of $\mathcal{I}_l$. 

Then, with  \textit{Assumption} \ref{ass:number_next_GC} the approximation of the optimal control law in (\ref{eq:Theta_k_next_3}) can be rewitten as
\begin{align}
\hat{\pi}_k^* &= \underset{\pi_k}{\arg\min} \sum_{i=1}^{N_k} \sum_{j=1}^{N_{targ}} \bigg\{\sum_{\imath \in \mathcal{I}^i_k} \bigg[  \mathcal{L}^{i,\imath}_k
\pi_k(i,\imath) 
+   \tilde{\mathcal{L}}^{\imath,j}_k \tilde{\pi}_k(\imath,j) \bigg]\bigg\}\nonumber\\
\text{s.t. } & \quad (\ref{eq:constraint_1}) - (\ref{eq:constraint_3}), \nonumber\\
& \quad \pi_k(i,\imath) = 0,  \; i = 1,\ldots,N_k \text{ and } \imath \notin \mathcal{I}^i_k, \nonumber\\
& \quad \tilde{\pi}_k(\imath,j) = 0, \, j = 1,\ldots,N_{targ} \text{ and } \imath \in \mathcal{I}_k^C
\label{eq:hat_pi_k_d_th}
\end{align} 
To solve the LP problem in (\ref{eq:hat_pi_k_d_th}), only $\vert \mathcal{I}_k \vert \times N_{targ}$ shortest-paths in the directed graph $\mathcal{DG}_k$ are required to calculate $\tilde{\mathcal{L}}_k^{\imath,j}$ at each time step, where ``$\vert \cdot \vert$" indicates the cardinality of a set.

\subsection{Interpolation of Robot PDFs in Sub-space of GMM} 
By using (\ref{eq:hat_pi_k_d_th}), the optimal control law and the corresponding control can be approximated based on the collocation Gaussian components. However, the distribution velocity defined in (\ref{eq:distribution_velocity_WG}) is affected by the setting of the collocation Gaussian components.
To remove the affect from the collocation Gaussian components, the distribution distance between two sequentially obtained robot PDFs is divided evenly by a user-defined distribution interval, $\bar{d}$, and more robot PDFs are interpolated between them.    

Let the approximate optimal control, $\wp_k^{goal} = \hat{\mathcal{C}}_k^*(\wp_k,m_k)$, denote the goal PDF at the $k$th time step, instead of the PDF at $(k+1)$th time step. The distribution distance, $d(\wp_k,\wp_k^{goal})$, is divided into $T_k = \lceil d(\wp_k,\wp_k^{goal})/\bar{d} \rceil$ segments, where ``$\lceil \cdot \rceil$" indicates the ceiling operator. Then, $(T_k-1)$ robot PDFs are interpolated between $\wp_k$ and $\wp_k^{goal}$ according to (\ref{eq:distribution_interpolation}), and the PDF $\wp_k^{goal}$ is treated as $\wp_{(k+T_k)}$. Therefore, the distribution velocity can be expressed by
\begin{equation}
v_{\tau}^{\wp} = \frac{d(\wp_k,\wp_k^{goal})}{T_k \cdot \Delta t} \approx \frac{\bar{d}}{\Delta t}\triangleq\bar{v}^{\wp}, \quad k\leq \tau \leq k+T_k
\label{eq:appro_distribution_velocity}
\end{equation} 
It means that the robot PDFs travel at a relatively smooth velocity. If $\bar{d} \ll d(\wp_k,\wp_k^{goal})$, the distribution velocity can be treated as a constant. 

\subsection{Optimal Microscopic Control Law}
\label{subsec:Microscopic_Control_Law}
Given the matrix of the robot microscopic states $\mathbf{X}_k = [\mathbf{x}_1(t_k),\ldots,\mathbf{x}_N(t_k)]$ at the $k$th time step and the approximated optimal control $\hat{\wp}^*_{k+1}$, the optimal microscopic control law can be determined by the artificial potential field method. The attractive potential is 
\begin{equation}
U^{attr}_k = 
\int_{\mathcal{W}} \big[\hat{\wp}^*_{k+1} - \gamma \tilde{\wp}_{k+1}(\mathbf{X}_k, \mathbf{U}_k))\big]^2(\mathbf{x}) d\mathbf{x}
\label{eq:distribution_potential}
\end{equation} 
where $\mathbf{U}_k = [\mathbf{u}_1(t_k),\ldots,\mathbf{u}_N(t_k)]$ is the matrix of the microscopic controls at the $k$th time step, and $\tilde{\wp}_{k+1}(\mathbf{X}_k, \mathbf{U}_k)$ is the estimated robot PDF at the $(k+1)$th time step given the microscopic states $\mathbf{X}_k$ and the microscopic controls $\mathbf{U}_k$. In this paper, the kernel density estimation (KDE) method is applied to estimate the PDF from microscopic robot states. Here, $0 < \gamma \leq 1$ is a scalar parameter to control the scattering strength of the robots. 
A larger $\gamma$ results in a more scattered distribution of robots. Meanwhile, for every individual robot, the repulsive forces from  the observed obstacles and the other robots are also considered. Let $\rho \left(\mathbf{x}_n(t_k),\mathbf{u}_n(t_k)\right)$ denote the shortest distance between the robot $\mathbf{x}_n(t_{k+1})$ and the obstacles or the other robots, given the control input $\mathbf{u}_n(t_k)$. The repulsive potential for the $n$th robot is defined by
\begin{equation}
U^{rep}_{k,n} (\rho) = \begin{cases}
\frac{1}{2} \left( \frac{1}{\rho} - \frac{1}{\rho_{rep}}\right)^2, & \text{ if } \rho \leq \rho_{rep} \\
0, & \text{otherwise} 
\end{cases}
\label{eq:repulsitive_potential}
\end{equation}
where $\rho_{rep}$ is the repulsive distance threshold to create a repulsion effect on the robot. The microscopic control inputs can be determined according to the sum of the attractive and repulsive gradients. More details can be found in \cite{ZhuGDM2019}. 

\section{Simulations and Results}
\label{sec:Simulations_Results}
In this section, a synthetic simulation is presented to evaluate the proposed ADOC approach, and the performance of the ADOC approach is compared with three state-of-the-art approaches in this section.

\subsection{Simulation Objectives and Settings}
The effectiveness of the ADOC approach presented in the previous sections is demonstrated on a network of $N=500$ mobile robots with single-integrator dynamics
\begin{equation}
\dot{\mathbf{x}}_n(t) = \mathbf{u}_n(t), \mathbf{x}_n(t_0) = \mathbf{x}_{n_0}, \quad n = 1,\ldots,N
\end{equation}
where $\mathbf{x}_n = [x_n,\,y_n]^T$ is the robot state, $x_n$ and $y_n$ are the robot $xy$-coordinates in the inertial frame, $\mathbf{u}_n = [u_n^x, u_n^y]^T$ is the microscopic control input, and $u_n^x$ and $u_n^y$ indicate the linear velocities in the $x$- and $y$-direction, respectively. Here, the noisy $\mathbf{w}(t)$ in (\ref{eq:dynamics}) is ignored for simplicity. 

\begin{figure}[htp]
	\centering
	\subfloat[]{\includegraphics[width=2.5in]{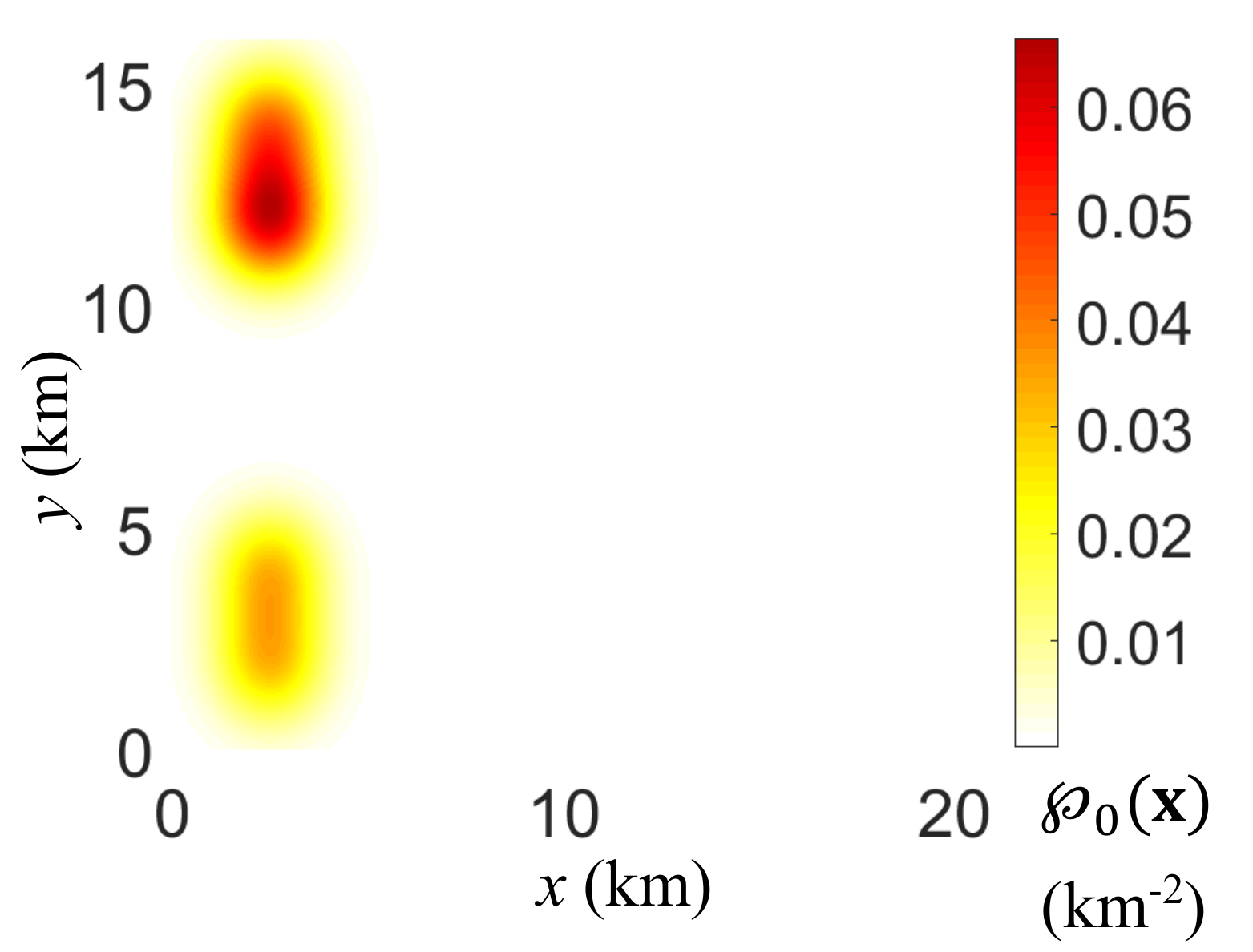}}
	\hfil
	\subfloat[]{\includegraphics[width=2.5in]{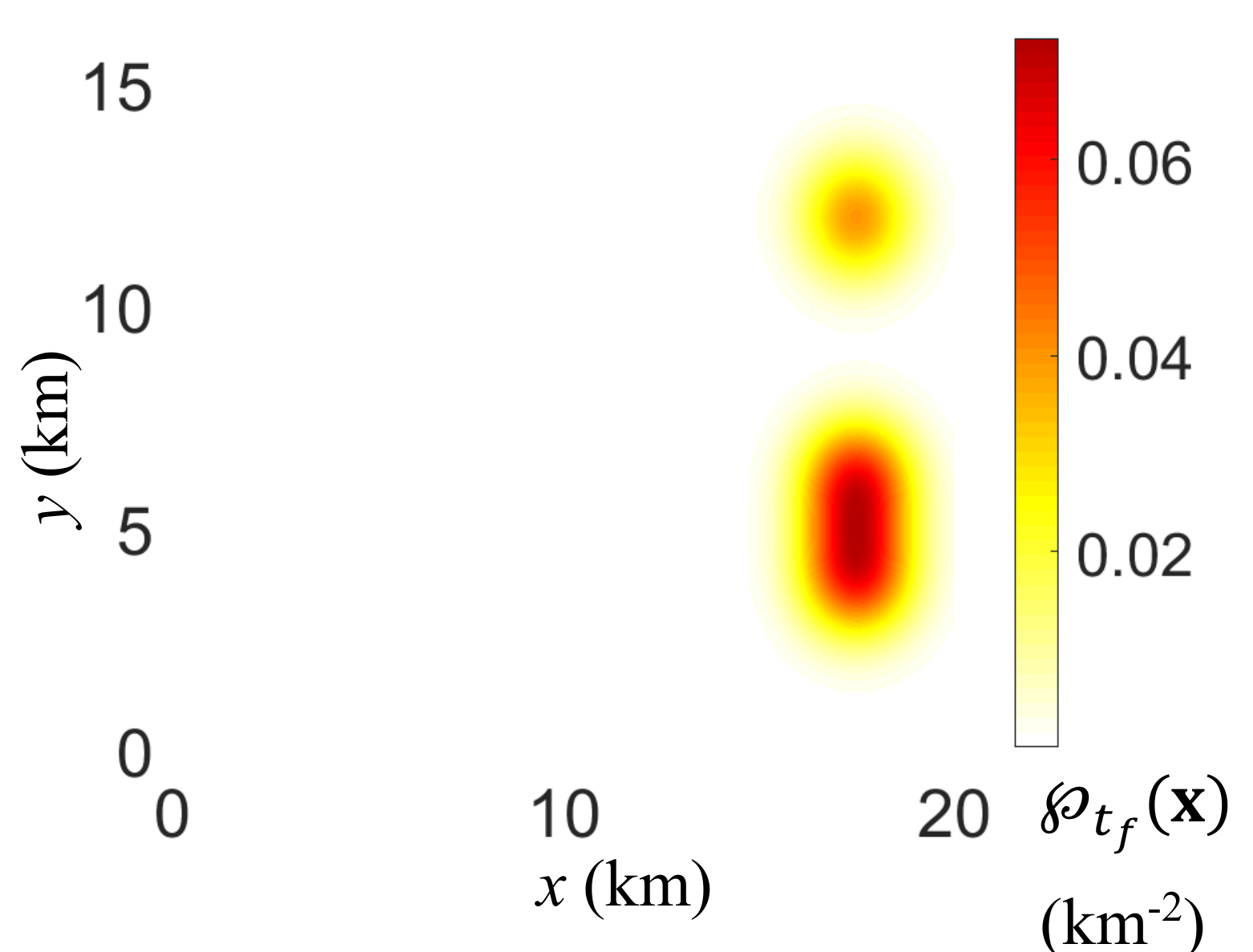}}
	\caption{Goal of VLSR system is to travel from an initial robot distribution shown in (a) to a goal robot distribution shown in (b).}
	\label{fig:Initial_Target_PDFs}
\end{figure}

\begin{figure}[htp]
	\centering
	\subfloat[]{\includegraphics[width=2.5in]{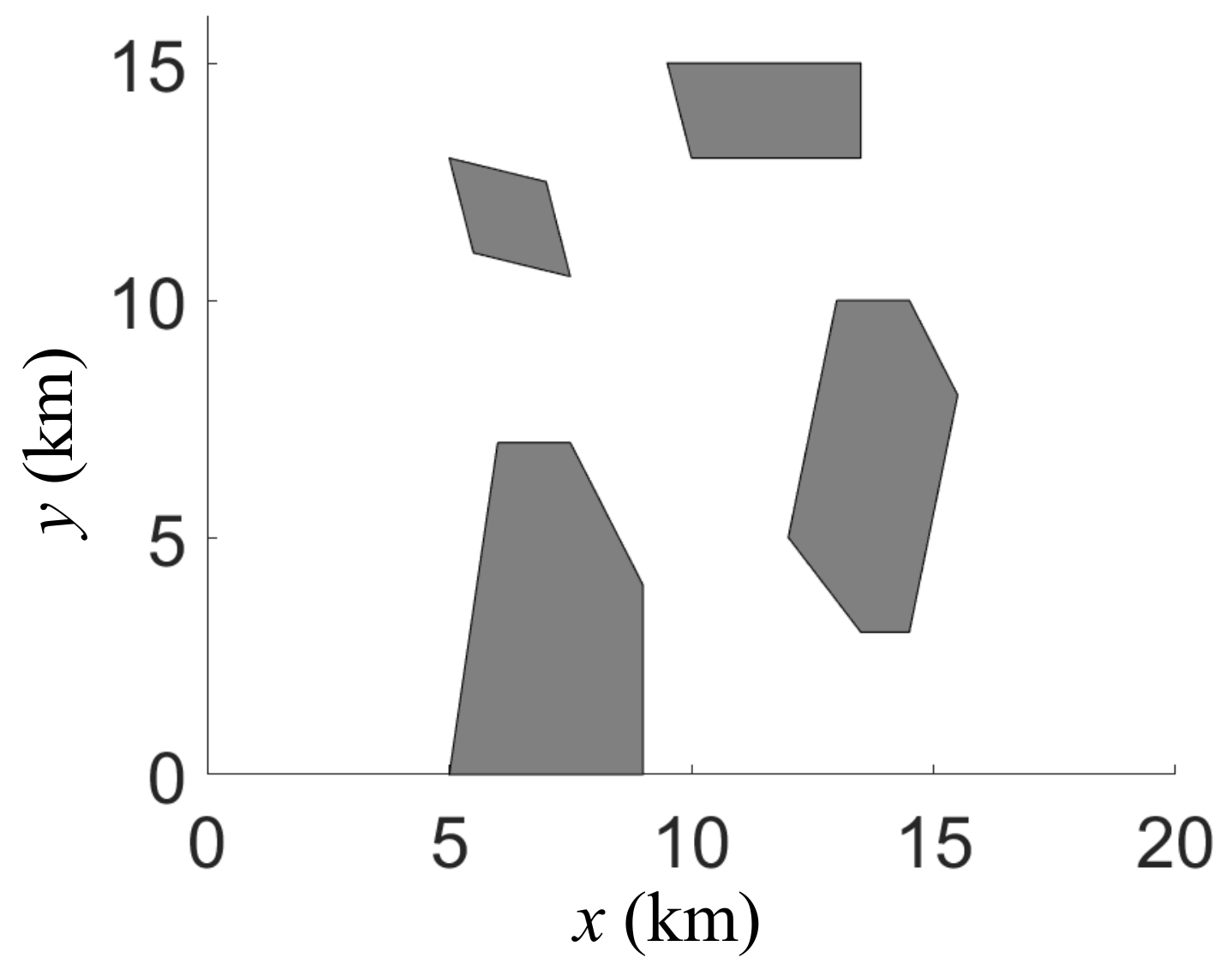}\label{fig:prioiri_layout}}
	\hfil
	\subfloat[]{\includegraphics[width=2.5in]{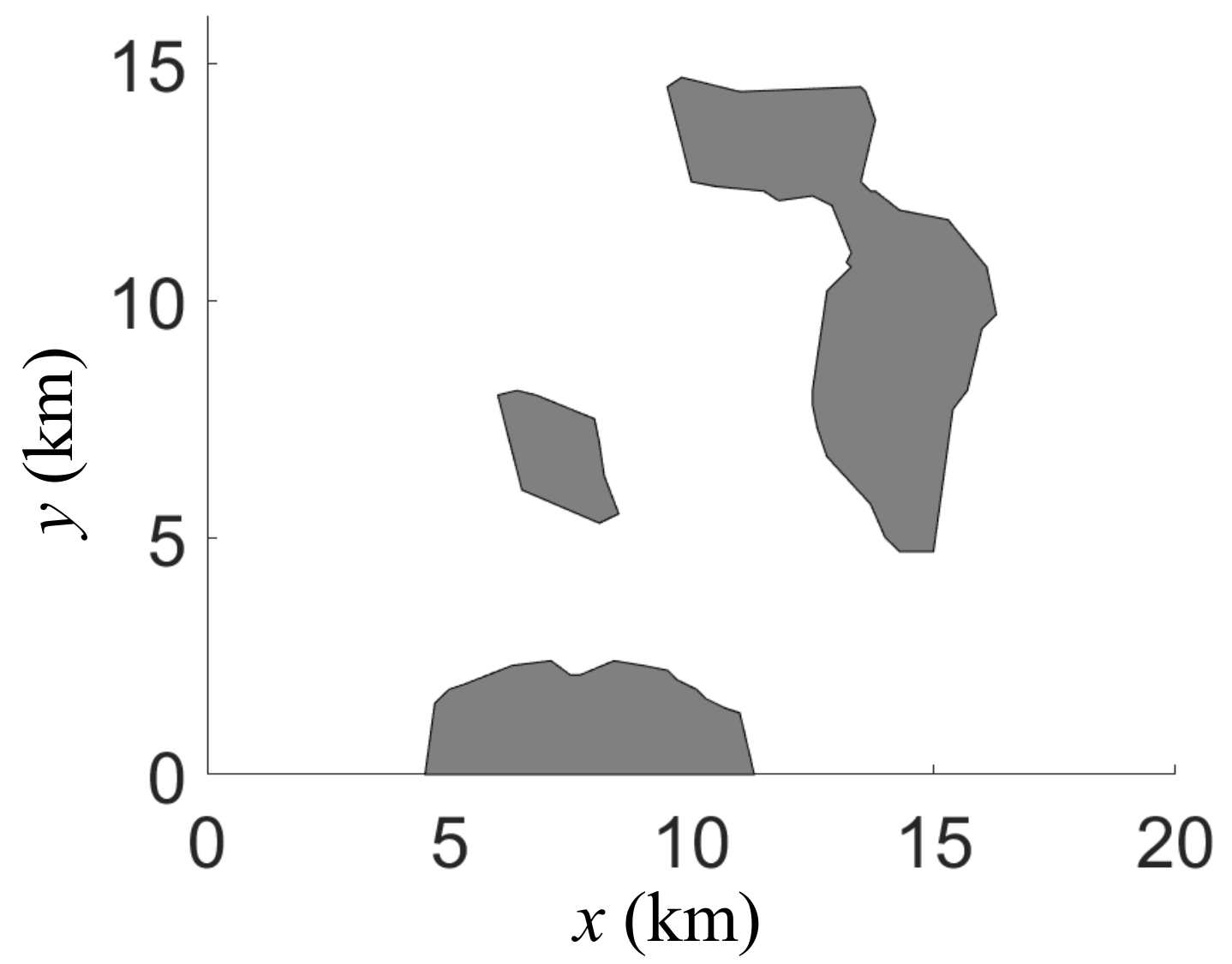}\label{fig:actual_layout}}
	\caption{Layout of obstacles. (a) A-priori layout of obstacles (b) Actual in situ obstacles }
	\label{fig:Priori_True_layout}
\end{figure}

As mentioned in Section \ref{sec:Problem_Formulation}, the objectives of the VLSR system is to travel from a given initial distribution $\wp_0 = \sum_{i=1}^{N_0}\omega_0^i g^i_0$ to a target distribution $\wp_{targ} = \sum_{j=1}^{N_{targ}} \omega^j_{targ} g^j_{targ}$ while avoiding collisions with the obstacles deployed in an ROI, $\mathcal{W} = [0,L_x] \times [0,L_y]$, where $N_0 = 4$, $N_{targ} = 3$, $L_x = 20$ km and $L_y = 16$ km. The initial and target robot distributions are presented in Fig. \ref{fig:Initial_Target_PDFs}. At the initial time, the a priori layout of obstacles is inaccurate (Fig. \ref{fig:prioiri_layout}). The actual layout of obstacles (Fig. \ref{fig:actual_layout}) is not available and needs to be observed and updated by the identical omnidirectional range sensors equipped on the mobile robots at every time step. In this simulation, the radius of the sensor FOV is $r = 1$ km. 

\subsection{Simulation of ADOC Approach}
For the ADOC approach, the observed and updated layout of obstacles at the $k$th time step is represented by the map function defined in (\ref{eq:map_function}), 
which is evaluated at the collocation points. These collocation points are deployed on the ROI $\mathcal{W}$ with the even spatial intervals, $\Delta x = \Delta y = 0.1$ km. Thus, there are $160 \times 200 = 32000$ collocation points in total. 

As described in \textit{Assumption} \ref{ass:collocation_GC}, the collocation Gaussian components, $\boldsymbol{G} = \{g^1_{c},
\ldots,g^{N_{c}}_{c}\}$, are set in the ROI $\mathcal{W}$, where $\mu^{\imath}_c = [\imath_x - 0.5, \, \imath_y - 0.5]$ km, $\imath_x = 1,\ldots,L_x$ and $\imath_y = 1,\ldots, L_y$, and $\Sigma_c = \begin{bmatrix}
0.5 & 0 \\ 0 & 0.5
\end{bmatrix}$ km\textsuperscript{2}. Thus, there are $N_c = 16 \times 20 = 320$ collocation Gaussian components deployed in the ROI and $N_c + N_{targ} = 320 + 3 = 323$ nodes in the directed graph, $\mathcal{DG}_k = (\boldsymbol{V}, \mathcal{E}_k)$, described in Section \ref{subsec:Approximation_in_SubSpace}. The distance threshold for the directed graph defined in \textit{Assumption} \ref{ass:number_next_GC} is set to $d_{th} = 4$ km. 

The fixed time interval is set to $\Delta t = 0.01$ hr. The user-defined distribution interval in (\ref{eq:appro_distribution_velocity}) is set to $\bar{d} = 0.05$ km to obtain a relatively constant distribution velocity, such that $\bar{v}^{\wp} = \frac{\bar{d}}{\Delta t} = \frac{0.05}{0.01} = 5$ km/hr. 

\begin{figure}[htp]
	\centering
	\subfloat[]{\includegraphics[width=2.5in]{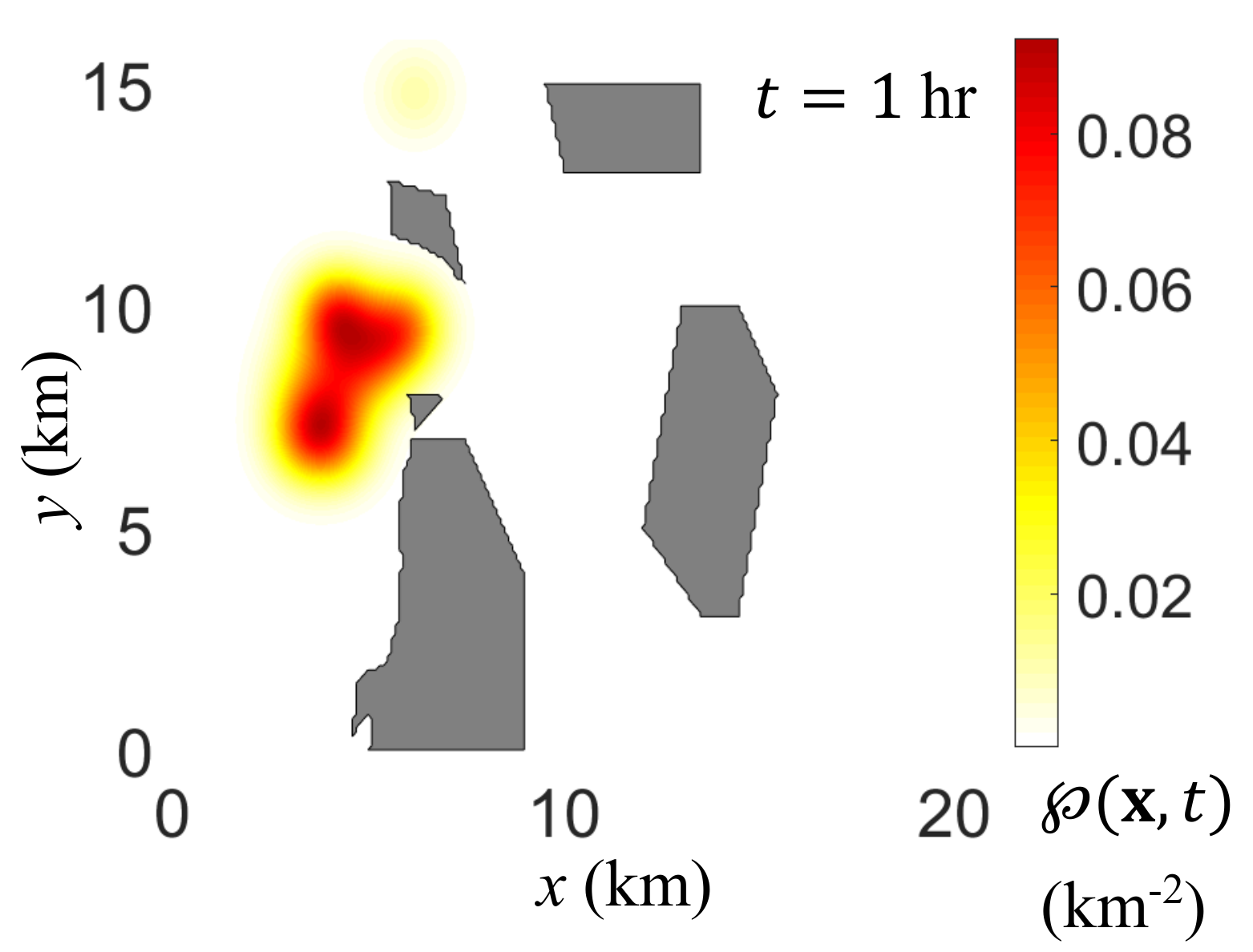}}
	\hfil
	\subfloat[]{\includegraphics[width=2.5in]{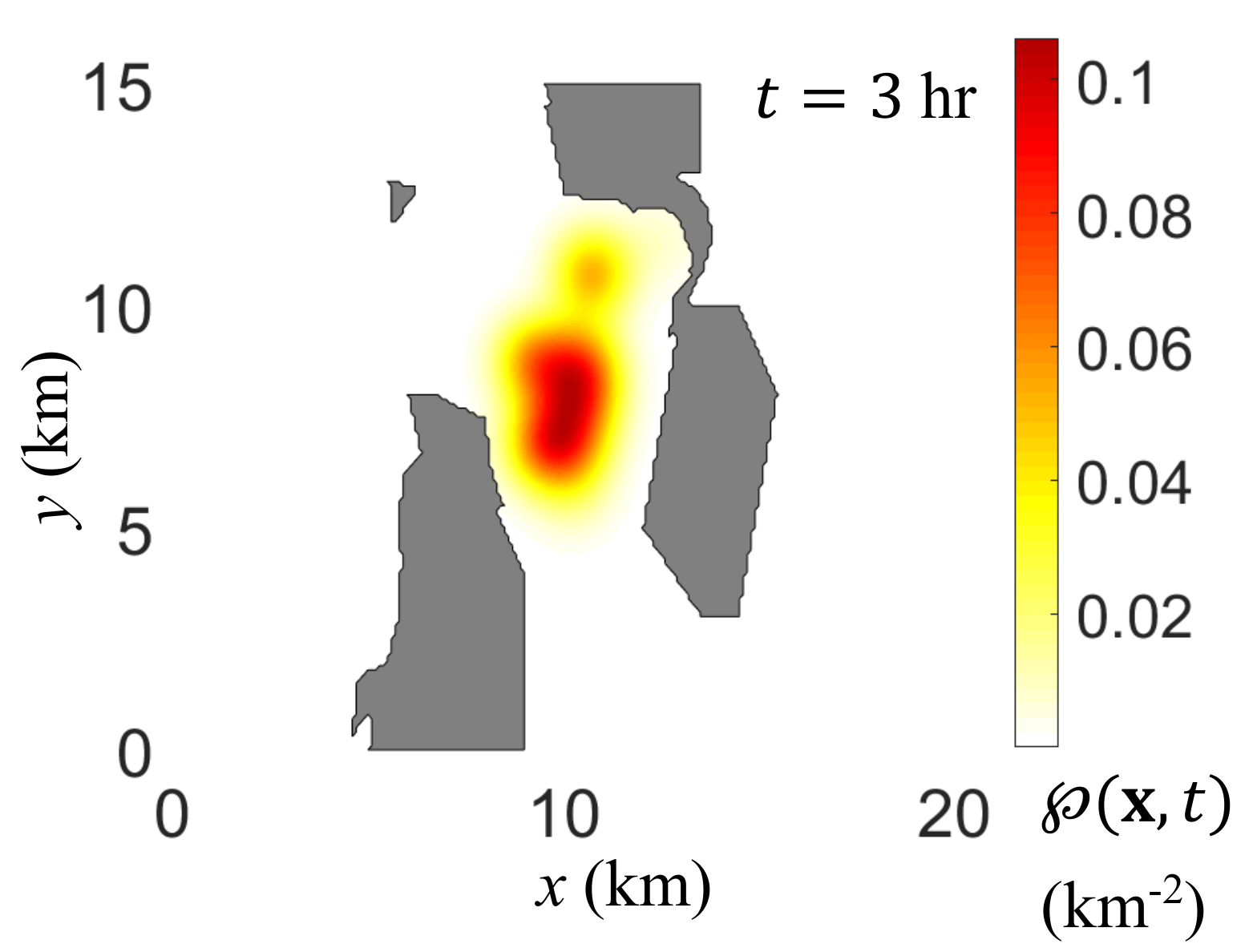} \label{fig:PDF_ADOC_300}}
	\hfil
	\subfloat[]{\includegraphics[width=2.5in]{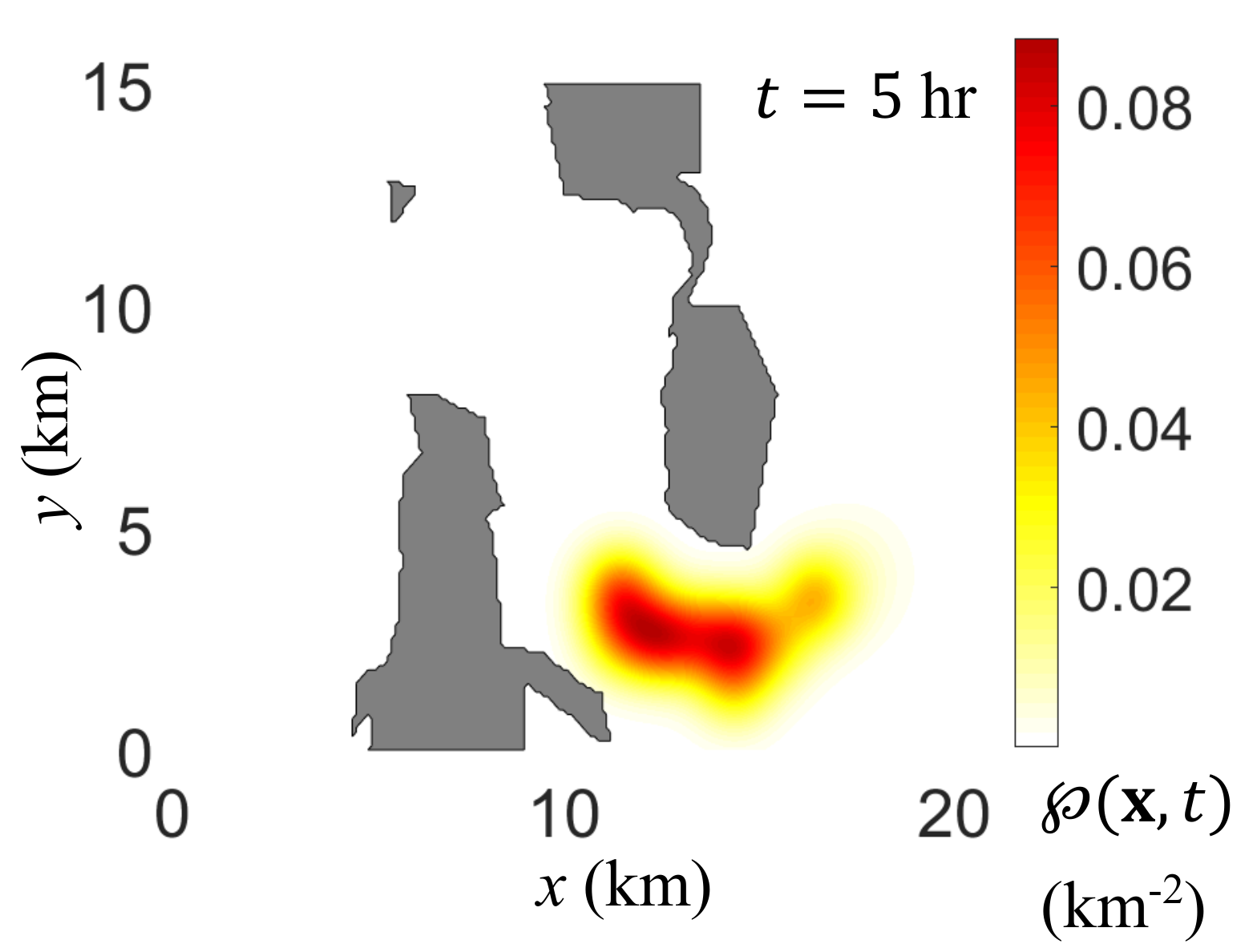}}
	\hfil
	\subfloat[]{\includegraphics[width=2.5in]{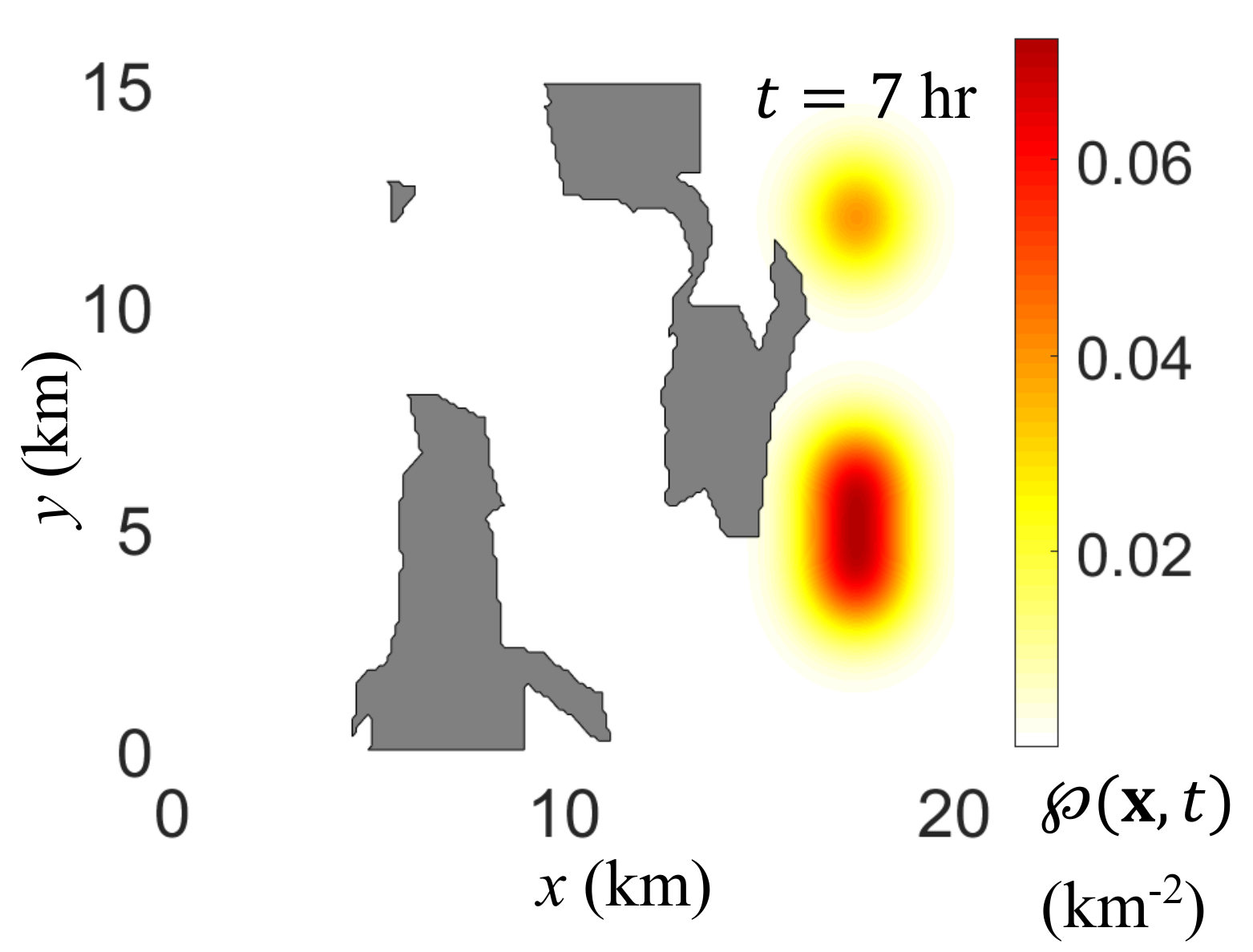}}
	\caption{Evolution of sensor PDFs generated by the ADOC approach at four moments, where the gray areas indicate the obstacles updated at each time step.}
	\label{fig:PDF_Trajectory_ADOC}
\end{figure}

\begin{figure*}[htp]
	\centering
	\subfloat[]{\includegraphics[width=2in]{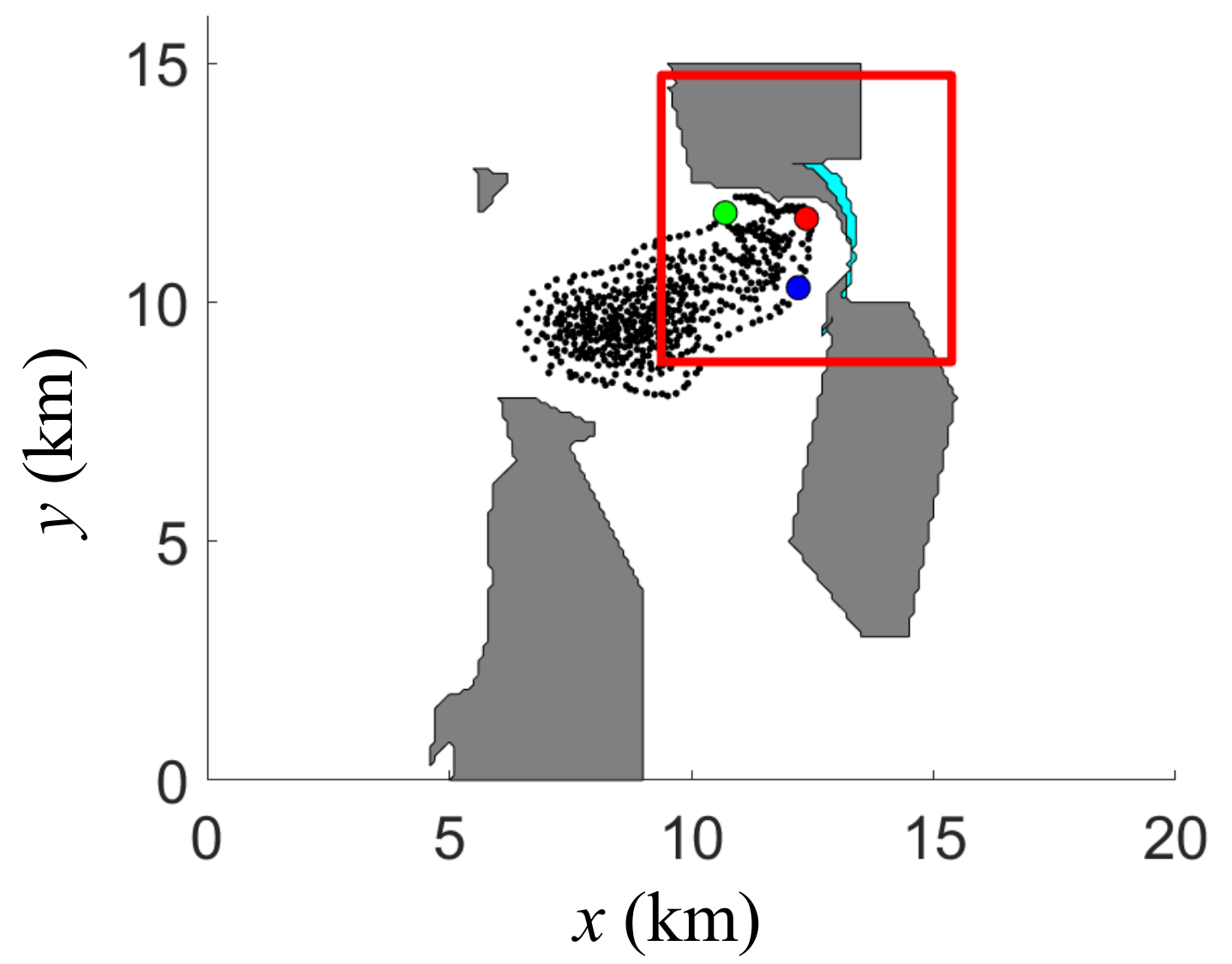}}
	\hfil
	\subfloat[]{\includegraphics[width=2in]{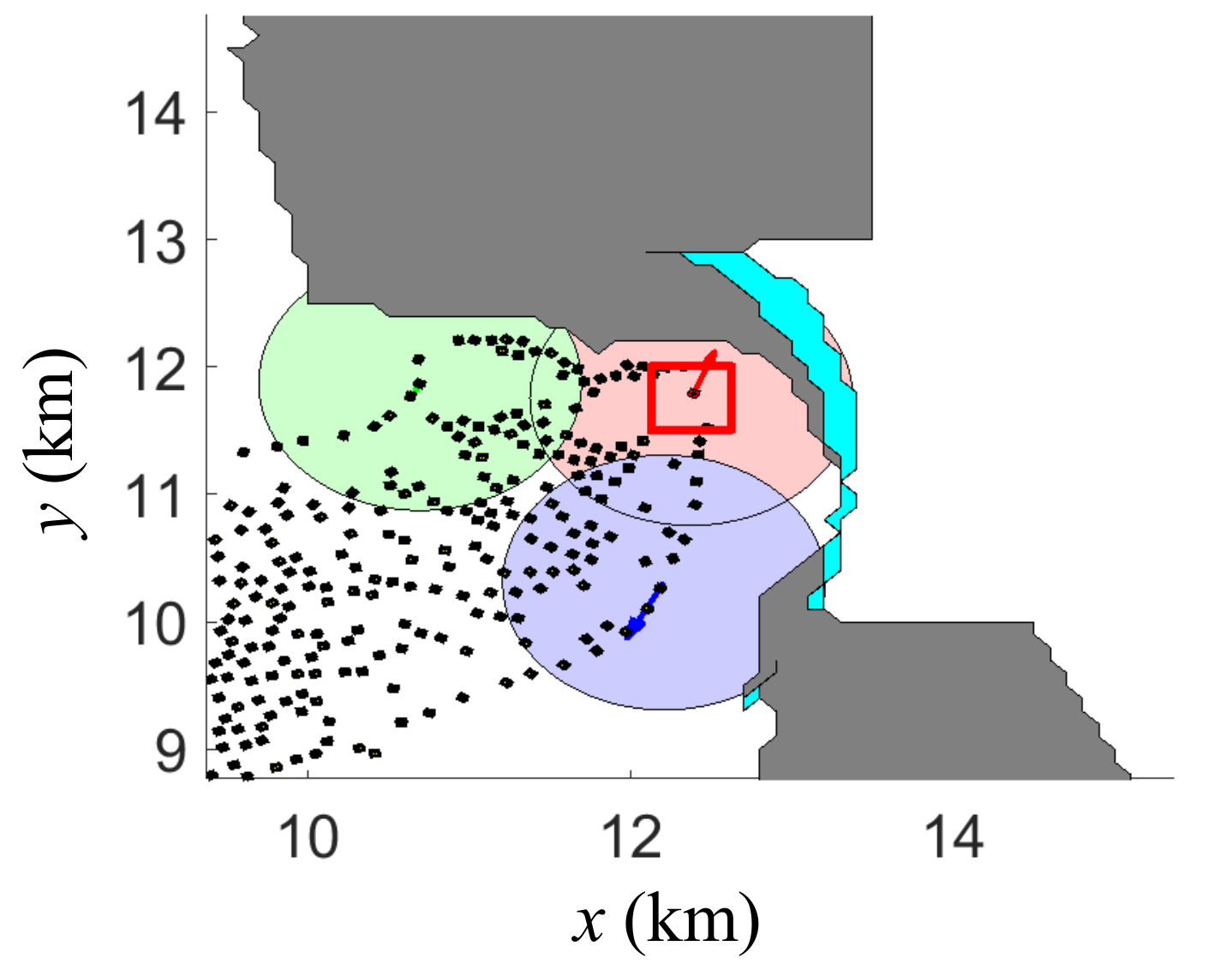}}
	\hfil
	\subfloat[]{\includegraphics[width=2in]{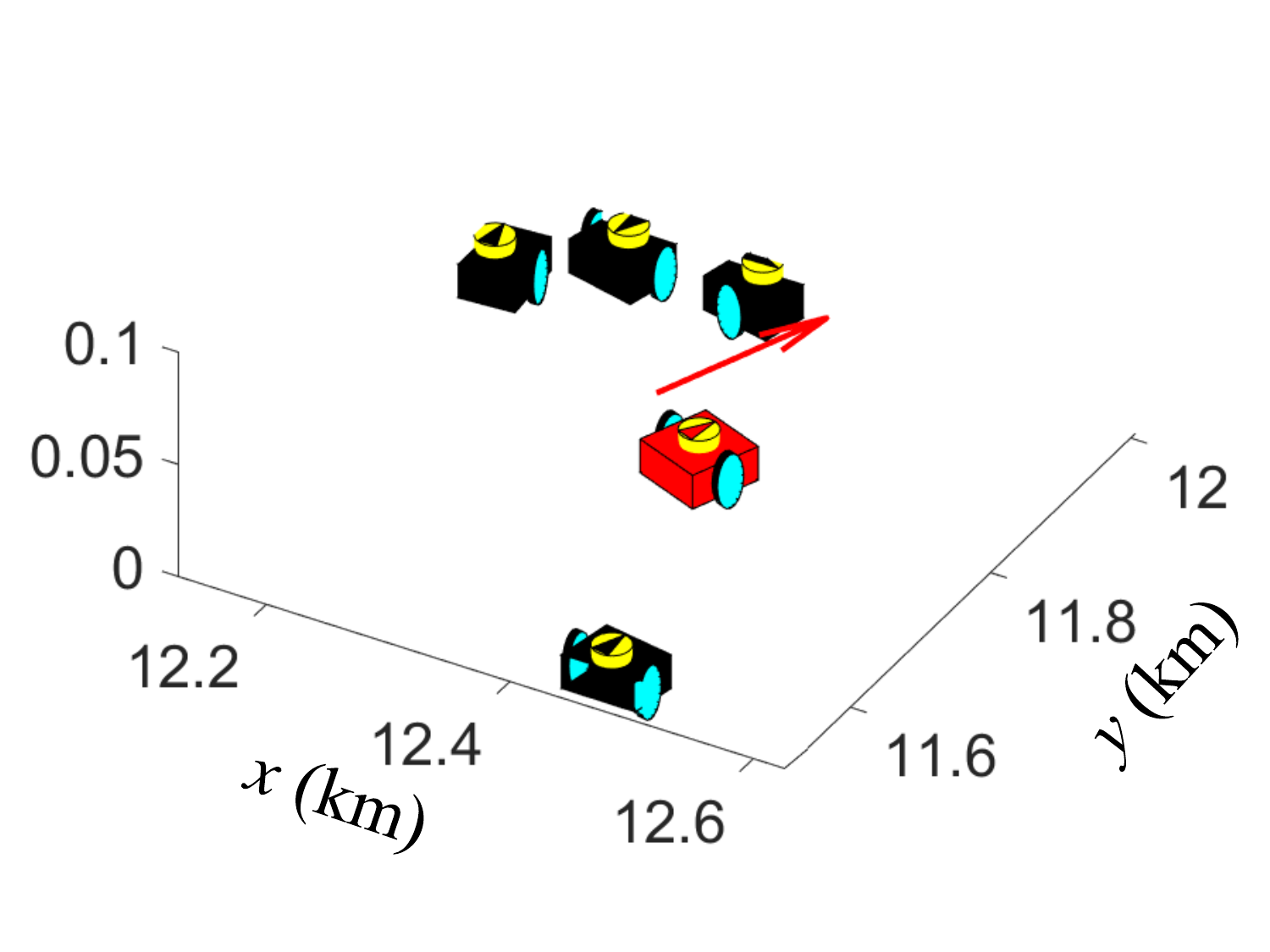}}
	\hfil
	\subfloat[]{\includegraphics[width=2in]{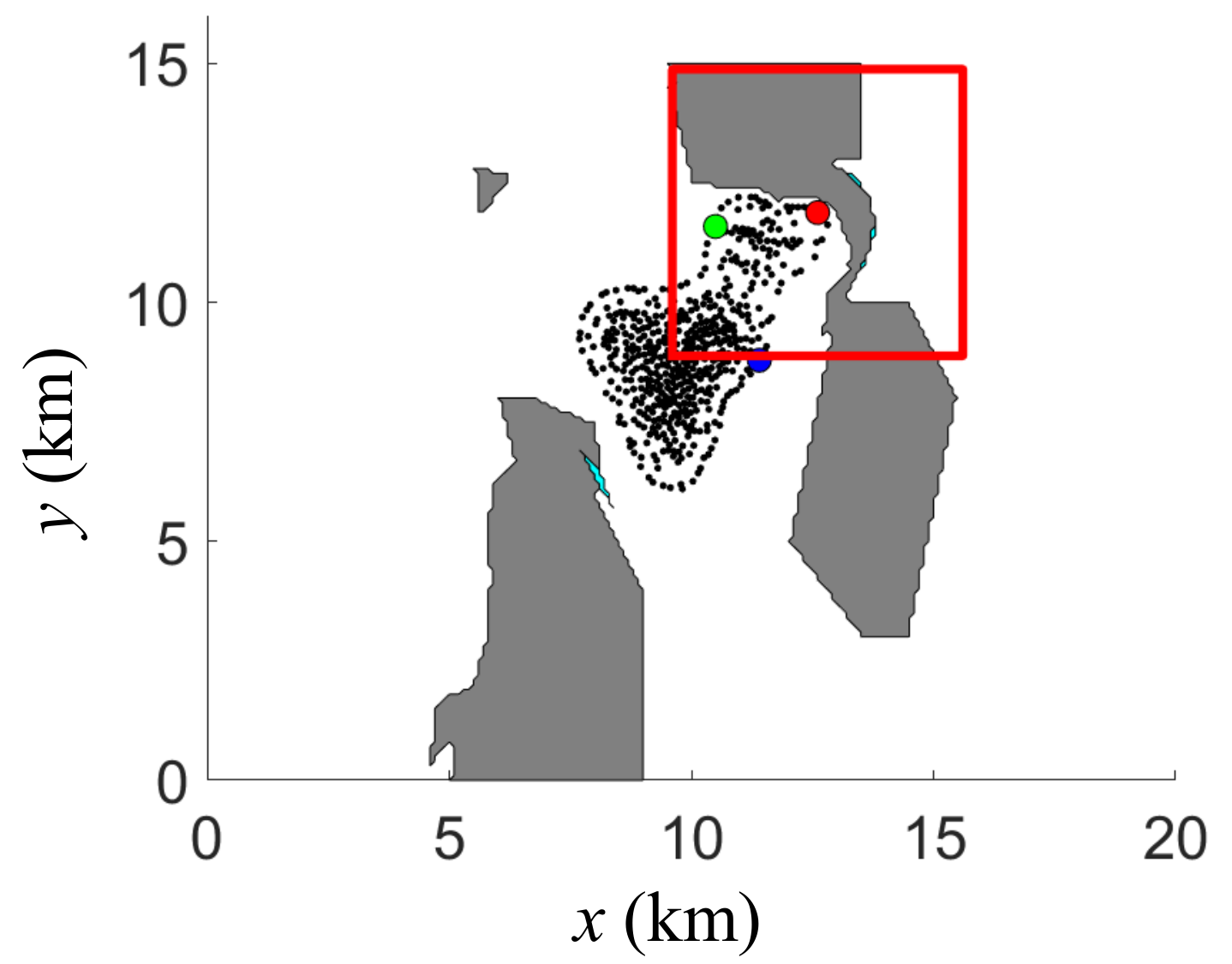}}
	\hfil
	\subfloat[]{\includegraphics[width=2in]{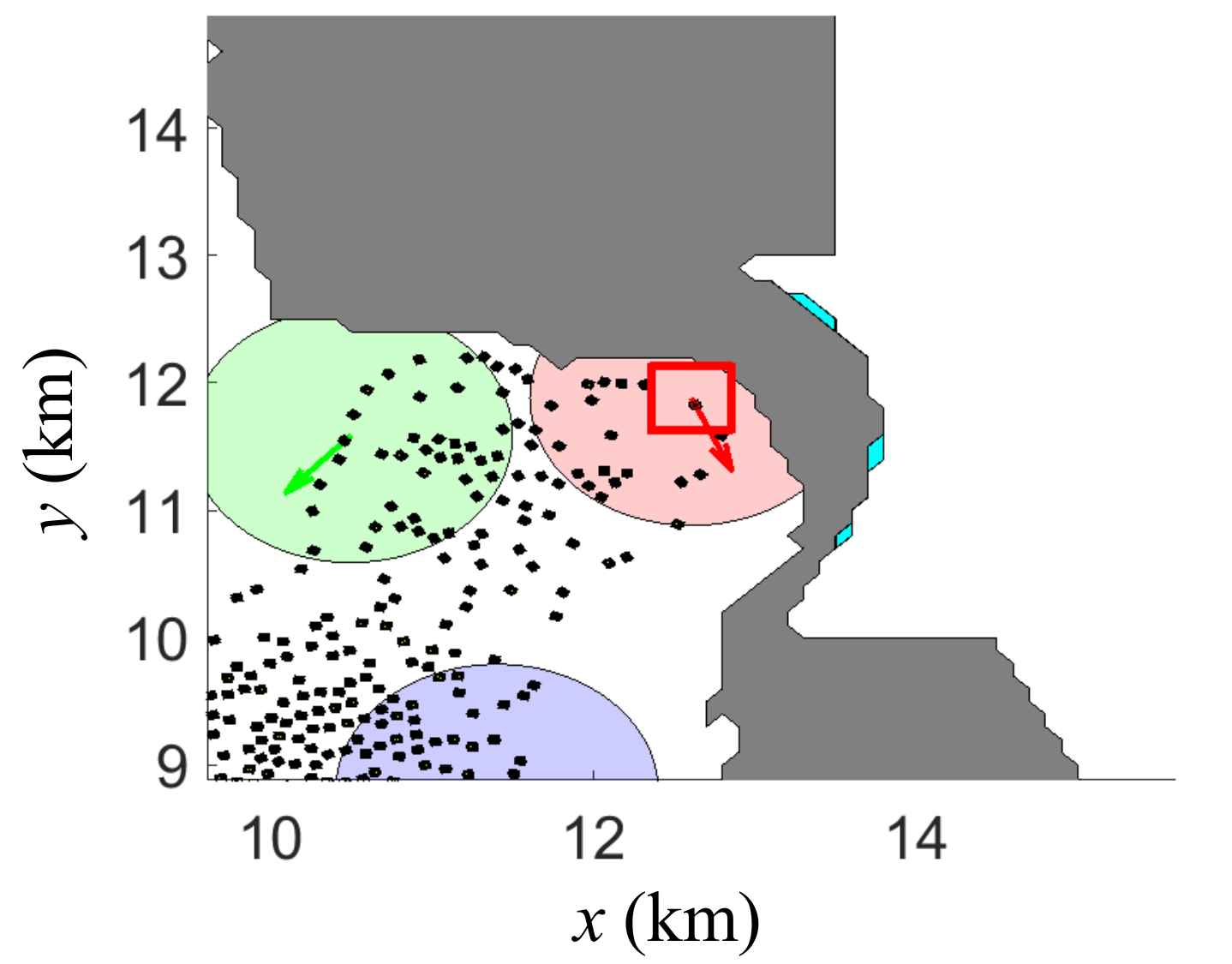}}
	\hfil
	\subfloat[]{\includegraphics[width=2in]{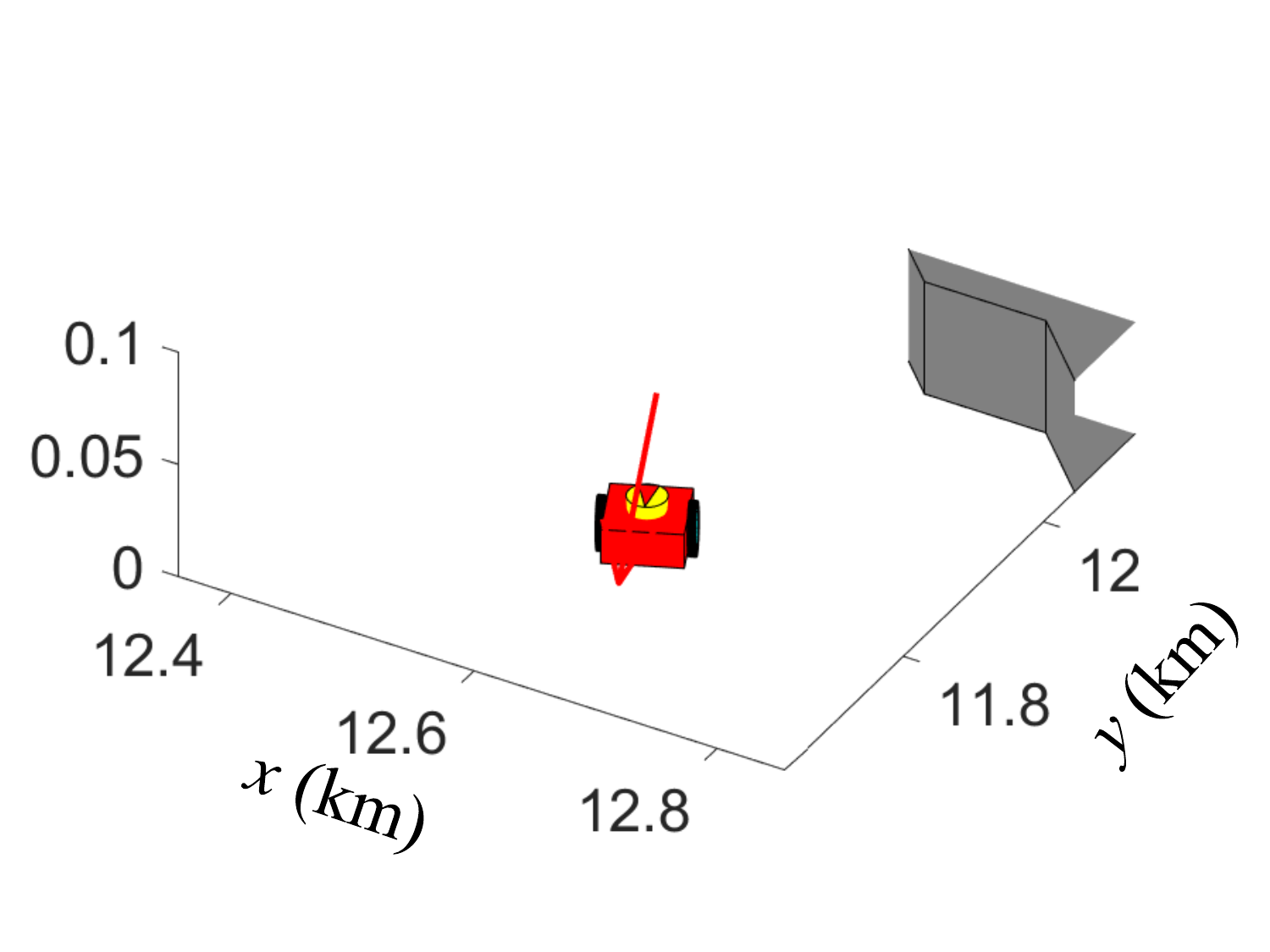}}
	\caption{ Positions and FOVs of robots at two different moments are presented in two rows, including $t_1=2.5$ hrs and $t_2=2.8$ hrs. The figures in the first column, (a) and (d), present the robot positions in the whole ROI. The figures in the second column, (b) and (e),are generated by zooming in the red bounding boxes in (a) and (d), respectively, where the FOVs of three robots are plotted in three different colors. The cyan areas indicate the obstacles which are observed during the past $0.1$ hrs. The figures in the last column, (c) and (f), are generated by zooming in the red bounding boxes in (b) and (e), respectively, where the robots and obstacles are plotted from a 3D viewpoint.}
	\label{fig:Trajectory_Take_Turning}
\end{figure*}

\begin{figure}[htp]
	\centering
	\includegraphics[width=3.2in]{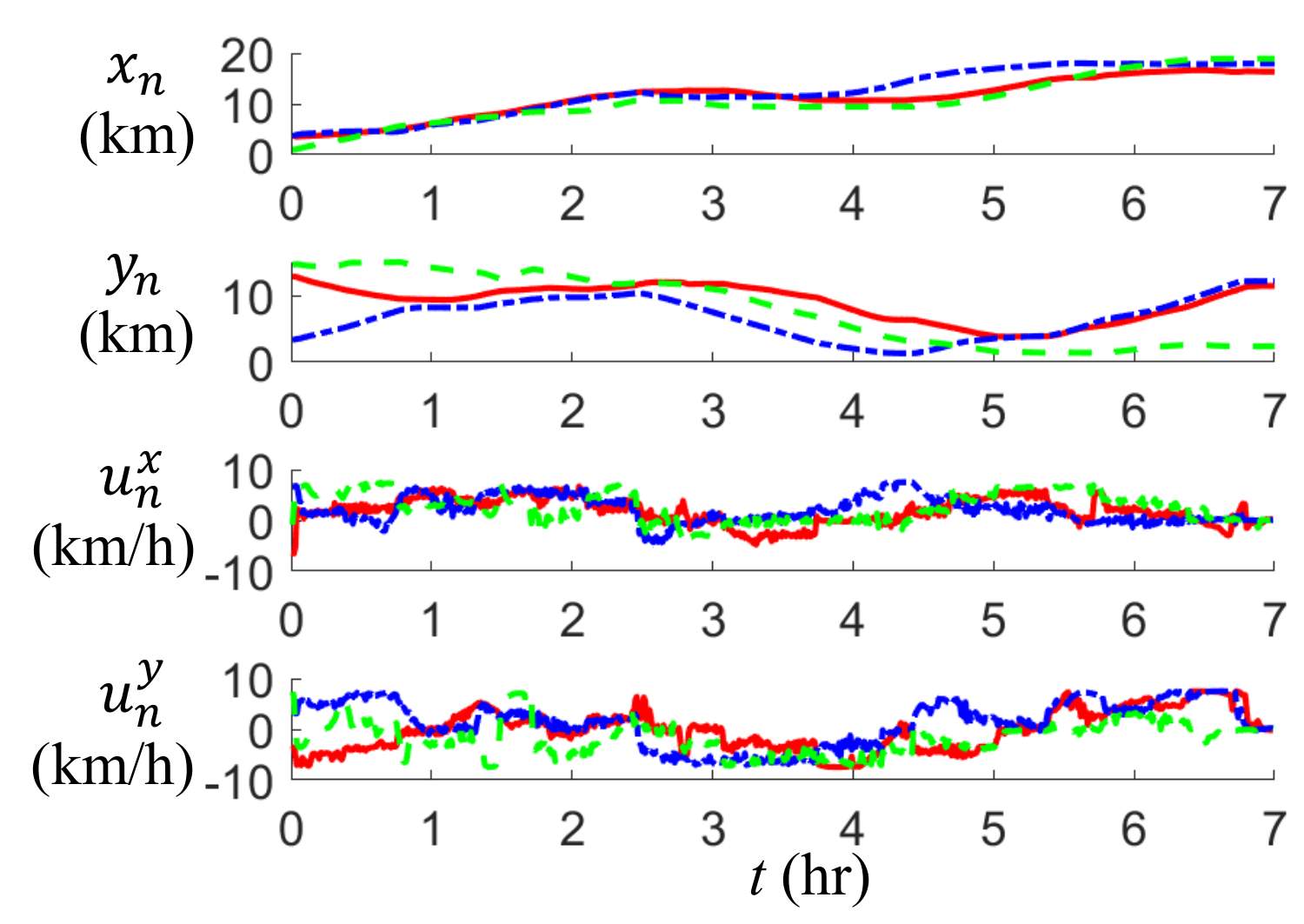}
	\caption{Microscopic state and control histories for three mobile robots chosen from the VLSR system.}
	\label{fig:Microscoic_state_control_history}
\end{figure}

For the microscopic control, the scalar parameter in (\ref{eq:distribution_potential}) is set to $\gamma = 0.85$. Furthermore, considering that the uncertainty of the observed obstacles, different repulsive distance thresholds are set to create the repulsion effects from the obstacles and the other robots, respectively, including the repulsive distance threshold for obstacles, $\rho_{rep}^{obs} = 0.3$ km, and the repulsive distance threshold for the other robots, $\rho_{rep}^{rob} = 0.1$ km.          
The proposed ADOC approach is applied to generate the approximated optimal macroscopic control law and the corresponding microscopic control inputs, which control the VLSR system to travel online in the ROI. The generated trajectory of robot PDFs are presented in Fig. \ref{fig:PDF_Trajectory_ADOC}, where the gray areas indicate the obstacles updated at each time step. From Fig. \ref{fig:PDF_ADOC_300}, it can be observed that at around $3$th hr, the new obstacle is observed and an original open path is blocked and the VLSR system can find new path adaptively to the target distribution. 
Positions and FOVs of robots at $t_1 = 2.5$ hr and $t_2 = 2.8$ hr are presented in Fig. \ref{fig:Trajectory_Take_Turning}, where the FOVs of three robots are plotted in three different colors. 
The microscopic state and control histories of these three robots are plotted in Fig. \ref{fig:Microscoic_state_control_history}. 
These results show that the VLSR system takes $T_f = 701$ steps and $T_f \Delta t = 701 \times 0.01 = 7.01$ hrs to successfully travel from the initial distribution to the target distribution under the control of the proposed ADOC approach.    

\subsection{Comparison of Performance}
To the best of our knowledge, there is no existing hoc ad approach for the problem described in Section \ref{sec:Problem_Formulation}, because of the number of robots is too large. Thus, three different benchmark approaches are developed based on the existing state-of-the-art techniques and compared with the proposed ADOC approach, including  probability-density-function-based artificial potential field (PDF-APF) approaches,
 Sampling-based artificial potential field (SAPF), and the sampling-based path-planning (SPP). 

\begin{itemize} 
	\item PDF-APF: Because the target PDF $\wp_{targ}$ is given, the attractive potential function in (\ref{eq:distribution_potential}) can be modified by
	\begin{equation}
	U^{attr}_k = 
	\int_{\mathcal{W}} \big[\wp_{targ} - \gamma \tilde{\wp}_{k+1}(\mathbf{X}_k, \mathbf{U}_k))\big]^2(\mathbf{x}) d\mathbf{x}
	\label{eq:distribution_potential_target_PDF}
	\end{equation}
	Meanwhile, consider the repulsive potential function with respect to the updated obstacles and robots in (\ref{eq:repulsitive_potential}). Then, the microscopic control inputs can be obtained according to the sum of the gradients of potential functions similar to Section \ref{subsec:Microscopic_Control_Law}.
	
	\item SAPF: Since the robot target PDF and the number of robots are both known, the set of the target robot positions $\mathcal{X}^{targ} = \{\mathbf{x}_n^{targ}\}_{n=1}^N$ can be obtained by sampling according to $\wp_{targ}$. Then, these target positions can be treated as attractive points generating the individual potentials to every robot, such that
	\begin{equation}
		U_{n,n^{\prime},k}^{attr} = -\Vert \mathbf{x}_{n}(t_k) - \mathbf{x}_{n^{\prime}}^{targ} \Vert^{-2}, \; n, n^{\prime} = 1,\ldots,N
	\end{equation} 
	which is the attractive potential from the $n^{\prime}$th attractive point to the $n$th robot. The attractive point exist if and only if the attractive position has not been occupied by any robot yet. Similarly, the repulsive potentials in (\ref{eq:repulsitive_potential}) are also considered to calculate the microscopic control inputs. 
    \item SPP: Like the SAPF approach, the target robot positions are obtained by sampling  according to $\wp_{targ}$. Then, the initial robot positions, $\mathcal{X}_0 = \{\mathbf{x}_n(t_0)\}_{n=1}^N$ and $\mathcal{X}^{targ}$ are paired according to their relative distances. Finally, the shortest-path-planing algorithm is applied with the time-varying map function $m_k$ to obtain the $N$ paths from the initial robot positions to the corresponding target positions while avoiding the collisions by using the repulsive potentials in (\ref{eq:repulsitive_potential}). 
\end{itemize}

Since there is no macroscopic state $\wp_k$ and the corresponding velocity $v^{\wp}_k$ for these compared approaches,  the microscopic velocities for robots are all set to $v^{rob} = 5$ km/hr. For the SPP approach, the shortest-path-planning algorithms are implemented based on a graph where the set of nodes is defined by $\mathbf{V}_{SPP} = \{\mu_c^{\imath}\}_{\imath = 1}^{N_c} \cup \mathcal{X}^{targ}$. For fairness reasons, the same parameters, \textit{e.g.}, $\rho_{rep}^{obs}$, $\rho_{rep}^{rob}$, and $d_{th}$, are applied in these approaches. Furthermore, although the time during $t_f$ is not required by these approaches, a maximum of the time steps is set to $T_f^{\max} = 2000$, which means that the algorithms will stop at the $T_f^{\max} = 2000$th time step even if the task is not completed.  

Among these approaches, ADOC and SPP approaches involve the path-planning algorithms, while all of the approaches use artificial potentiates to obtain the microscopic controls, including the attractive and repulsive potentials. Given the number nodes $N_{node}$ and number of edges $N_{edge}$ in a directed graph, the computational complexity of planning a shortest path from a one single source to a single target is $O\big(N_{edge} + N_{node} \log\log(N_{node})\big)$ according to \cite{ThorupShortestPathPlanning2004}. Moreover, the calculations of artificial potentials between attractive or repulsive points and every robots are all considered. Let $N_k^{obs}$ denote the number of collocation points occupied by obstacles up to the $k$th time step. Thus, the computational complexities at the $k$th time step of these four approaches are tabulated in TABLE \ref{tab:computational_complexity}. It is noteworthy that for the microscopic control, since $N_k \ll N$ and $N_{targ} \ll N$, the computational complexities for all approaches are the same. Furthermore, for the case where $N_k \cdot \vert \mathcal{I}_k \vert \cdot N_{targ} < N$, the proposed ADOC approach outperforms the SPP approach in the aspect of computational complexity. 
\begin{table*}
	\caption{Comparison of Computational Complexity}
	\label{tab:computational_complexity}
	\centering
	\begin{tabular}{c|c|c}
		\hline
		Approach & Path Planning & Microscopic Control  \\
		\hline
		ADOC     & 	$O\bigg( \big( \vert \mathcal{E}_k \vert  + \vert \boldsymbol{V} \vert  \log \log \vert \boldsymbol{V} \vert \big) N_k \vert \mathcal{I}_k\vert N_{targ} \bigg)$			 &			$O(NN_k + N N_k^{obs} + N^2)$			\\
		\hline	 
		PDF-APF &		N/A		 &	$O(NN_{targ} + N N_k^{obs} + N^2)$					\\
		\hline
		SAPF	&       N/A      &  $O(N N_k^{obs} + N^2)$      \\
		\hline
		SPP	    &    $O\bigg( \big( \vert \mathcal{E}_k^{SPP} \vert  + \vert \mathbf{V}^{SPP} \vert  \log \log \vert \mathbf{V}^{SPP} \vert \big) N \bigg)$ & $O(N N_k^{obs} + N^2)$\\
		\hline
	\end{tabular}
\end{table*}   

\begin{figure}[htp]
	\centering
	\subfloat[]{\includegraphics[width=2.5in]{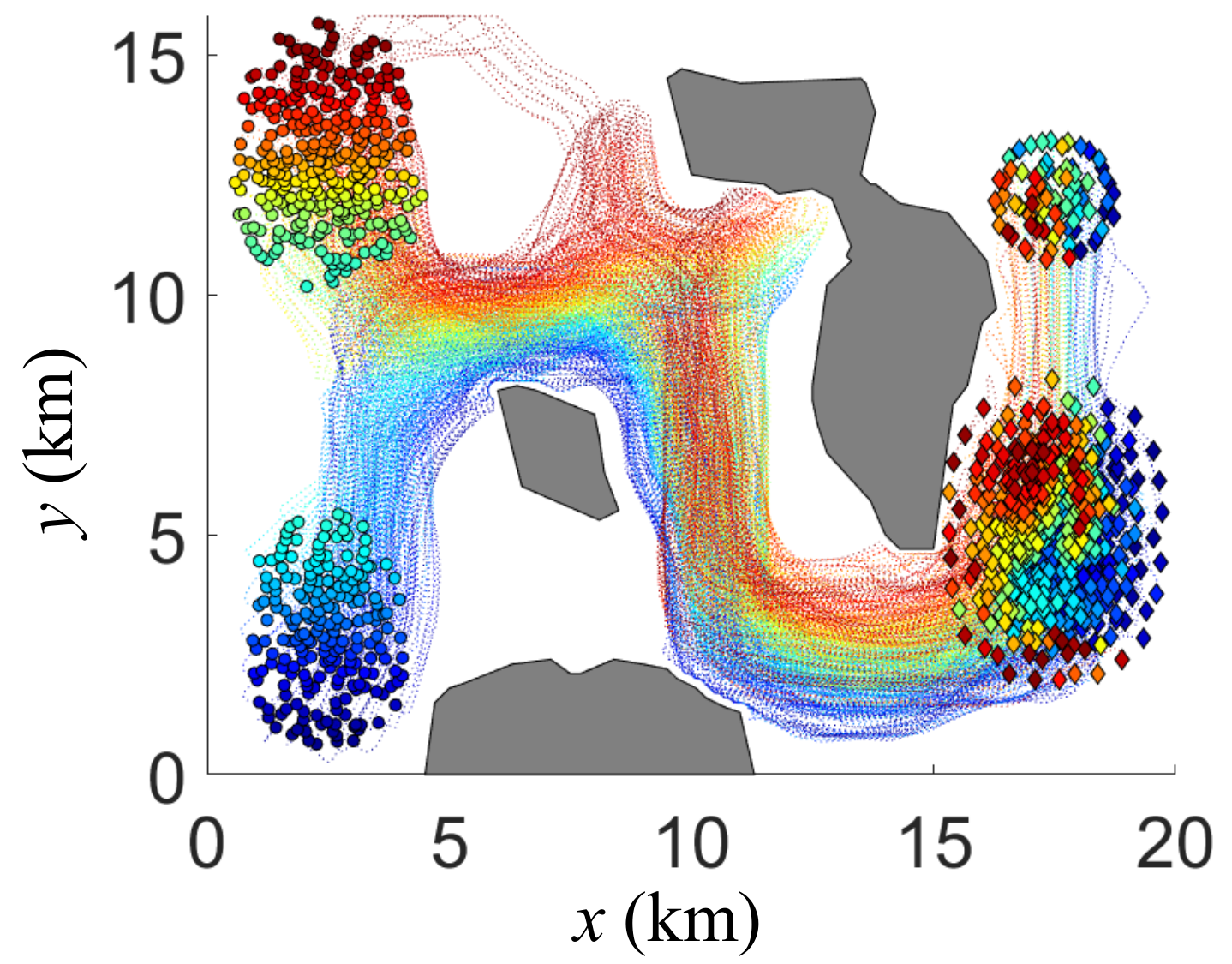}}
	\hfil
	\subfloat[]{\includegraphics[width=2.5in]{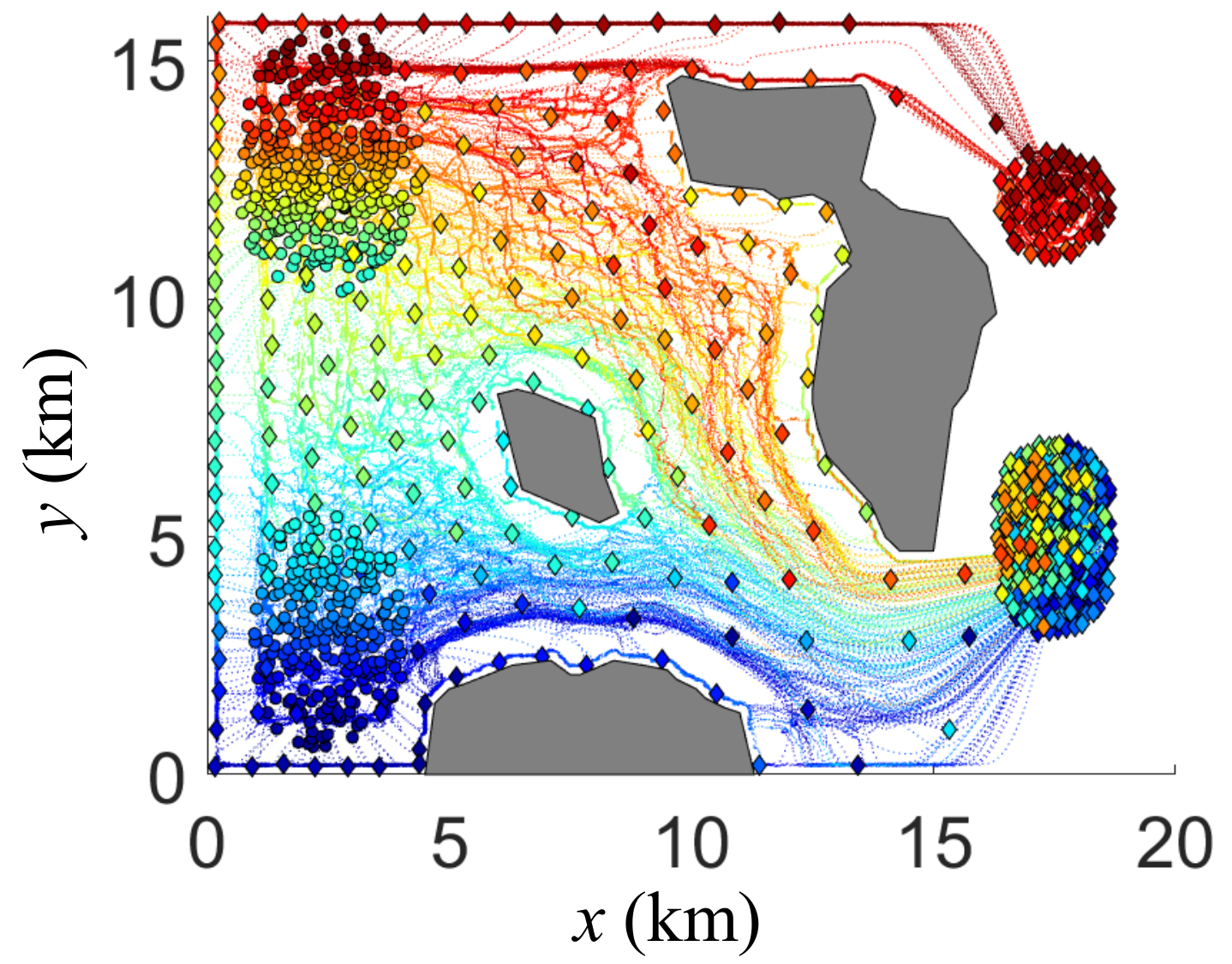}} 
	\hfil
	\subfloat[]{\includegraphics[width=2.5in]{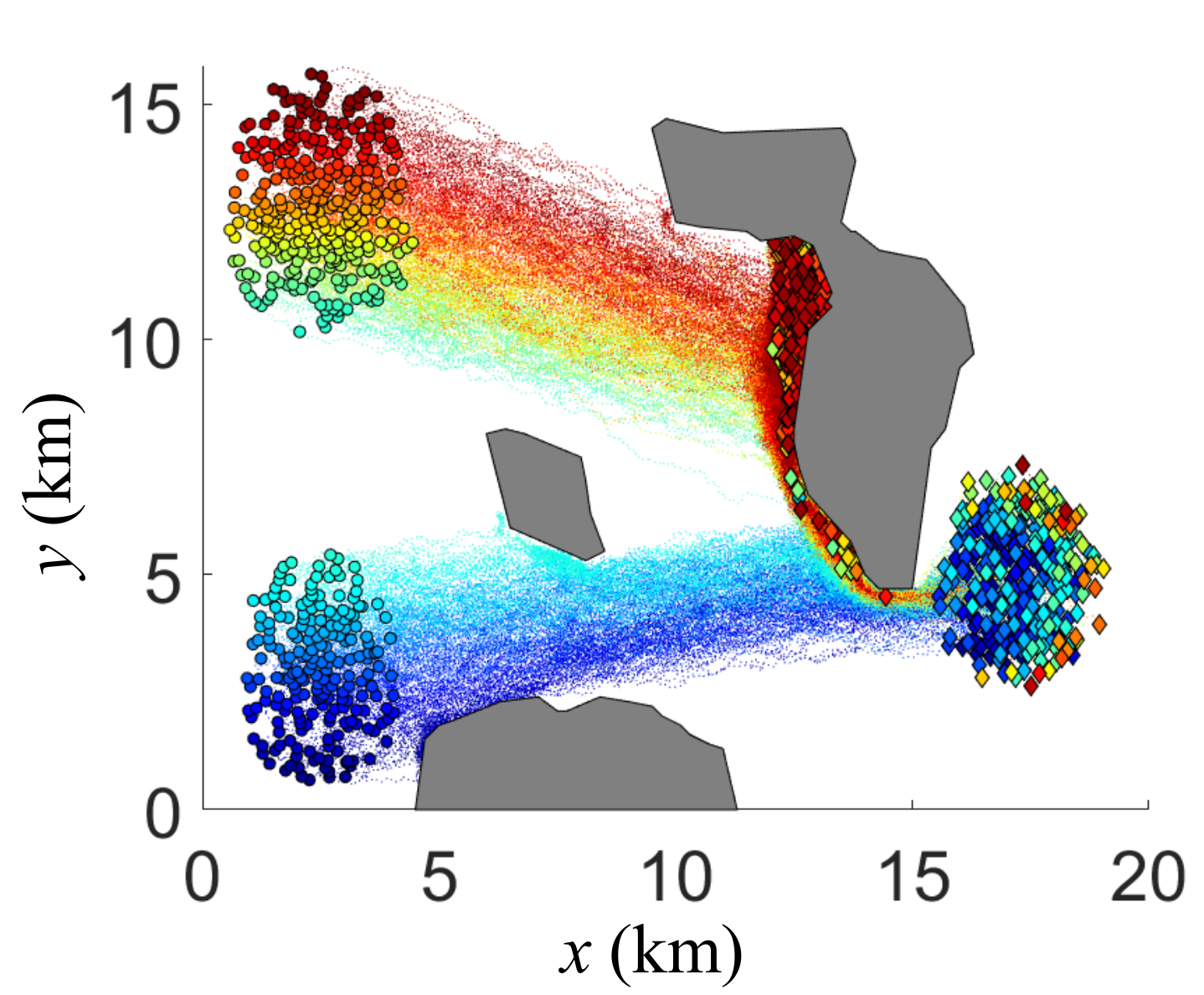}}
	\hfil
	\subfloat[]{\includegraphics[width=2.5in]{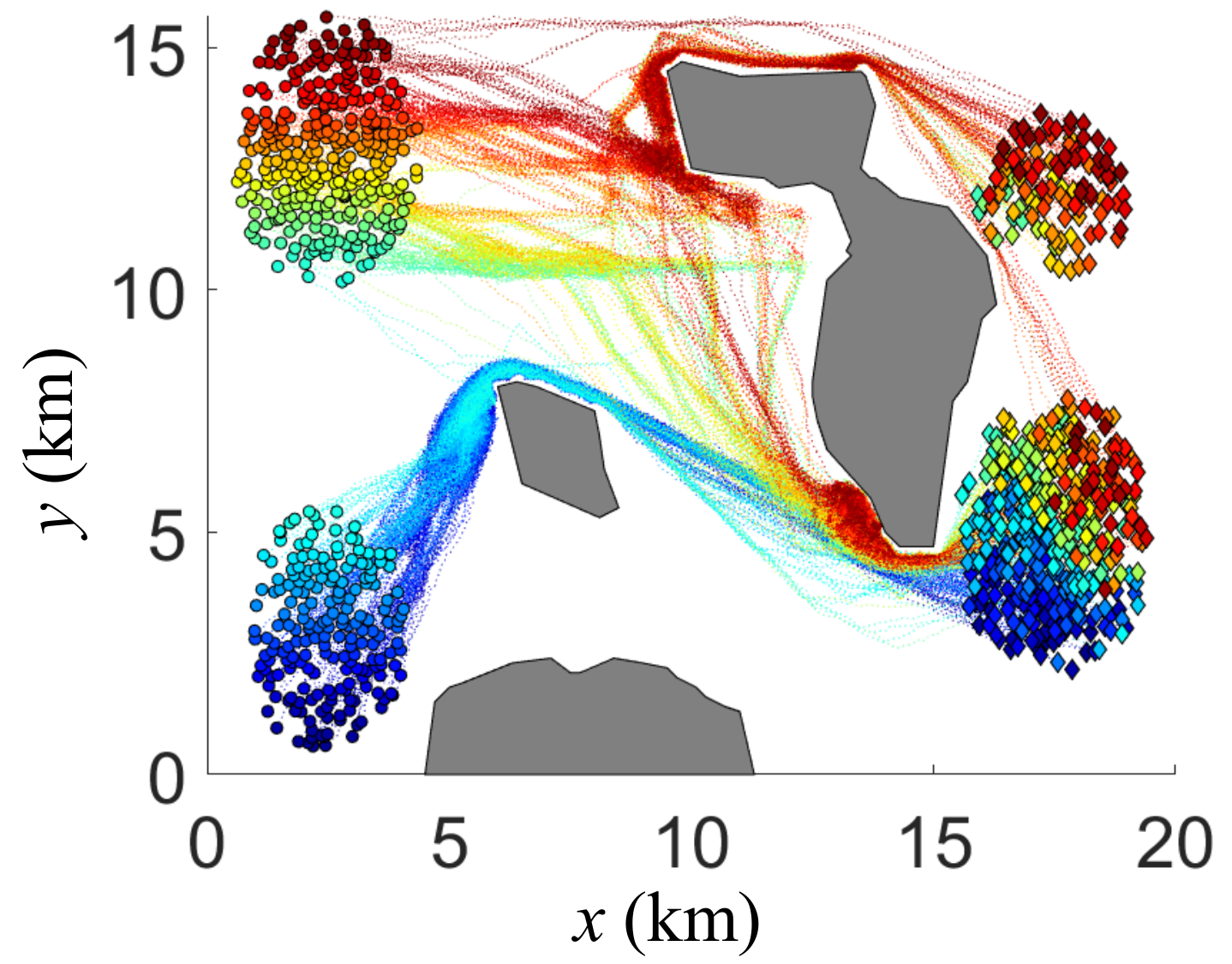}}
	\caption{Trajectories of robots generated by the different four approaches, including (a) ADOC, (b) PDF-APF, (c) SAPF, and (d) SPP, where the initial and final positions of robots are indicated by circles and diamonds, respectively, and the gray areas indicate the true obstacles in the workspace.}
	\label{fig:Robot_Trajectories}
\end{figure}

\begin{figure}[htp]
	\centering
	\subfloat[]{\includegraphics[width=2.5in]{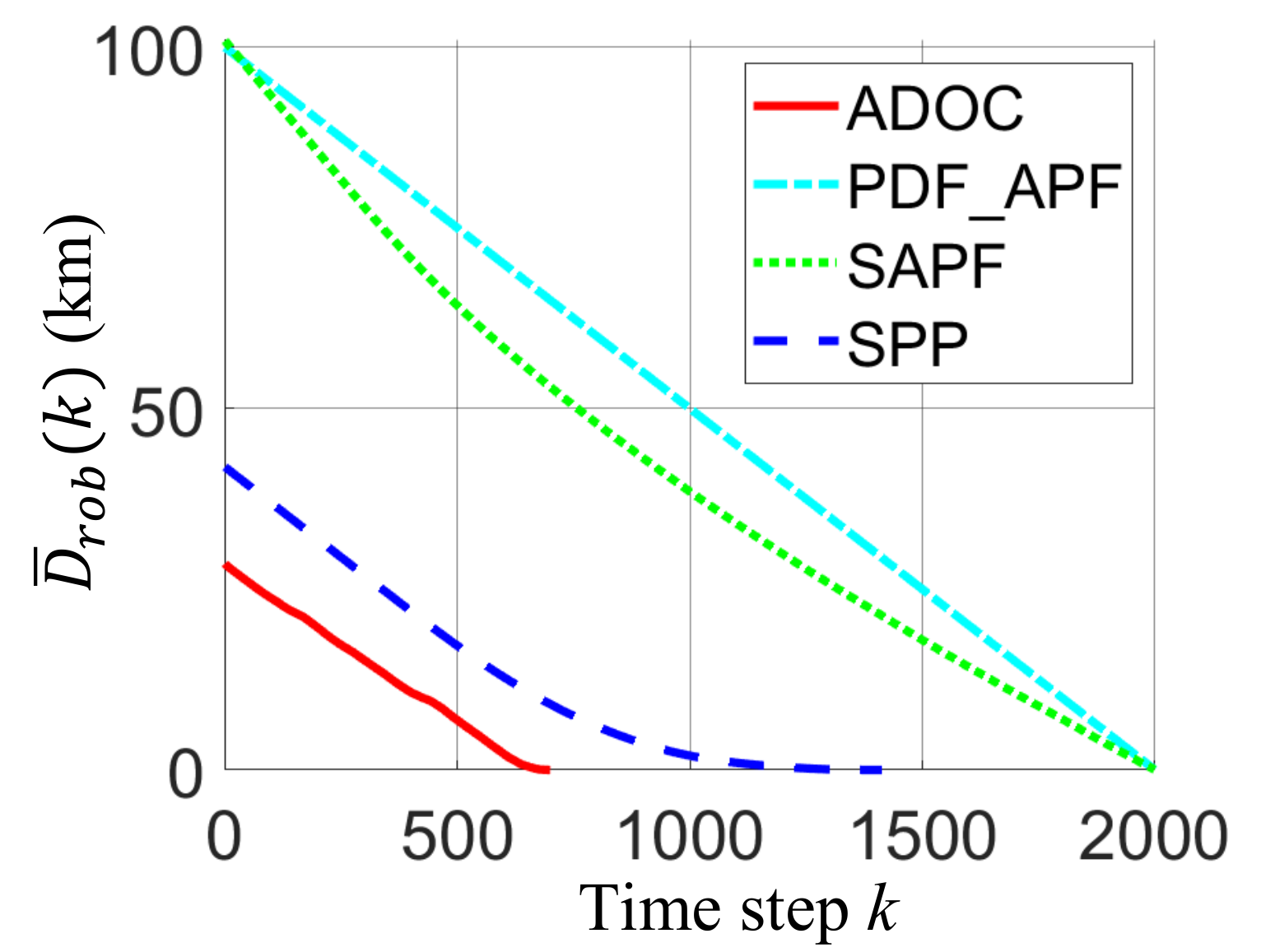}}
	\hfil
	\subfloat[]{\includegraphics[width=2.5in]{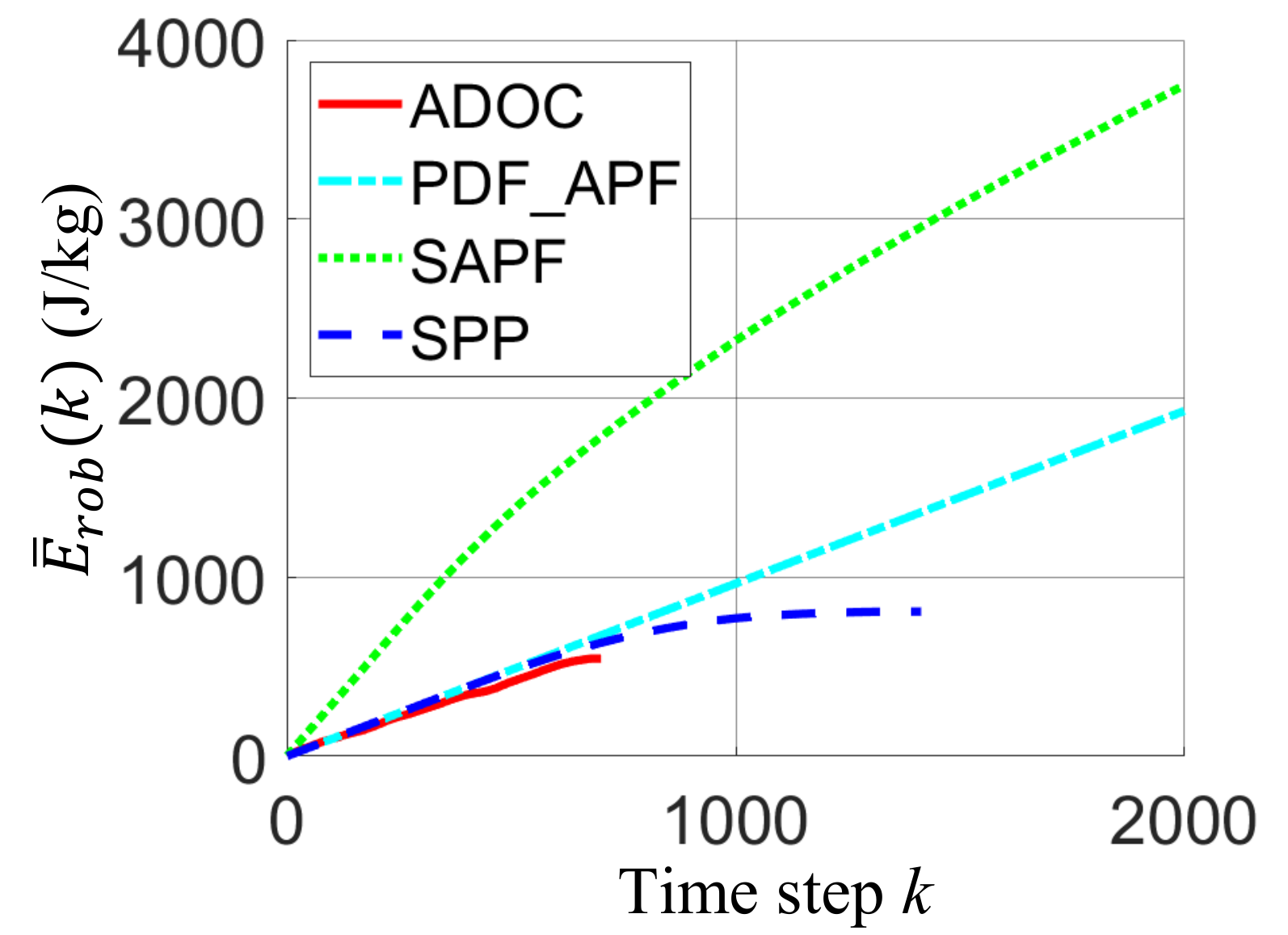}} 
	\caption{Performances of the VLSR systems generated by four approaches. (a) the average distance-to-go and (b) the engery-cost-per-kg.}
	\label{fig:Robot_performances}
\end{figure}

The trajectories of robots are plotted in Fig. \ref{fig:Robot_Trajectories}, which are generated by the different  approaches, including ADOC, PDF-APF, SAPF, and SPP. It can be observed that the ADOC and the SPP approaches complete the task successfully, while the PDF-APF and the SAPF approaches fail within $T_f^{\max} = 2000$ time steps. More numerical performances of these approaches are tabulated in TABLE \ref{tab:numerical_performances}, including the total time steps, $T_f$, the running time, the average distance-to-go $\bar{D}_{rob}(k)$, and the average energy-cost per kg, $\bar{E}_{rob}(k)$, which are defined by
\begin{equation}
\bar{D}_{rob}(k) = \frac{1}{N} \sum_{n=1}^N \sum_{\tau = k}^{T_f-1} \Vert   \mathbf{x}_n\big((\tau+1)\Delta t\big) - \mathbf{x}_n(\tau \Delta t) \Vert 
\end{equation}

\begin{equation}
\bar{E}_{rob}(k) = \frac{\eta}{2N} \sum_{n=1}^N \sum_{\tau = 1}^{k} \left[\frac{ \Vert \mathbf{x}_n(\tau \Delta t) - \mathbf{x}_n\big((\tau-1)\Delta t\big) \Vert }{\Delta t} \right]^2
\end{equation}
where $\eta = (1000/3600)^2 $ is applied for unit conversion. Here, these approaches run on the same PC with 18 cores. 
The curves of the distance-to-go and the energy-cost-per-kg are  plotted in Fig. \ref{fig:Robot_performances}. 
These results show that the proposed ADOC approach complete the task more efficiently and effectively than the other approaches. 

\begin{table}
	\caption{Comparison of Numerical Performances}
	\label{tab:numerical_performances}
	\centering
	\begin{tabular}{c|c|c|c|c}
		\hline
		         &       &                       &                 & \\
		Approach & $T_f$ & Running Time & $\bar{D}_{rob}(0) $ & $\bar{E}_{rob}(T_f)$  \\
		         &       & (min)               &  (km)    &  (J/kg)      \\
		\hline
		ADOC     & 	 \textbf{701} &	\textbf{37.0544} &  \textbf{28.5623}   & \textbf{545.1525}		\\
		\hline	 
		PDF-APF & 	2000 & 60.2447 &  99.95    &	1928.0478 	\\
		\hline
		SAPF	& 	2000 &   19.6193 &  100.8671   &  3747.0543   \\
		\hline
		SPP	    & 	1414 &  141.2573 &  41.9044   &  808.1635   \\
		\hline
	\end{tabular}
\end{table}

\section{Conclusion and Future Research}
In this paper, an ADOC approach is proposed to carry out online cooperative sensing and navigation tasks for the VLSR systems in highly uncertain environments, which is an extension of the DOC approach, where the VLSR system is described by the robot PDFs referred to as the macroscopic state. Because This approach is formulated as an RL-ADP problem in the Wasserstein-GMM space based on the optimal mass transport (OMT) theorem, it can also be considered as an online MARL approach. To the best of our knowledge, this is the first implementation of online MARL for continuous states and controls, which provide a novel research direction of online MARL approaches. 

The proposed ADOC is a centralized approach where the optimal control law and the corresponding microscopic controls are all generated based on observations from all robots. Considering that the GMM is a linear combination of Gaussian components, it is feasible to divide the generation process of optimal control law into several groups according to the Gaussian components. Then, a decentralized version of the ADOC approach can be proposed in future research and can be applied to more general applications, \textit{e.g.}, the problems in swarm robotics.


%

\appendices
\section{Lower bound of optimal control law}
\label{Appedix:lower_bound_of_optimal_control_law}
\begin{IEEEproof}
First, consider the case of $k^{\prime} = k-1$, $0 < k <T_f$ and $k \leq l < T_f$. From any robot PDF $\wp_l$,  
by using the optimal functional control law obtained at the $k^{\prime}$th time step,  $\mathcal{C}_{k^{\prime}}^*$, recursively, a trajectory of robot PDFs $\{\wp_{\tau}\}_{\tau = l}^{T_f}$ are generated associated with the obstacle map function $m_k$.

In addition, consider the optimal value functional $\mathcal{V}_k^*(\wp_l,m_k)$, $k \leq l < T_f$. According to (\ref{eq:VF_m_k}) and the Bellman equation,  $\mathcal{V}_k^*(\wp_l,m_k)$ can be expressed by
\begin{align}
    \mathcal{V}_k^*(\wp_l,m_k) &= \min_{\wp_{l+1}} \left[ \mathscr{L}(\wp_l,m_k,\wp_{l+1})+ \mathcal{V}_k^*(\wp_{l+1},m_k) \right] \nonumber\\
    &= \mathscr{L}(\wp_l,m_k,\wp_{l+1}^*) + \mathcal{V}_k^*(\wp_{l+1}^*,m_k) \nonumber\\
    &\leq \mathscr{L}(\wp_l,m_k,\wp_{l+1}) + \mathcal{V}_k^*(\wp_{l+1},m_k)
    \label{eq:Bellman_equation_mk}
\end{align}
where $\wp_{l+1} = \mathcal{C}^*_{k^\prime}(\wp_l,m_k)$ and $\wp^{*}_{l+1} = \mathcal{C}^*_k(\wp_l,m_k)$. 

By recursively utilizing  (\ref{eq:Bellman_equation_mk}), the following inequality is obtained
\begin{align}
    \mathcal{V}_k^*(\wp_l,m_k) &\leq \mathscr{L}(\wp_l,m_k,\wp_{l+1}) + \mathcal{V}_k^*(\wp_{l+1},m_k) \nonumber\\
    &\leq \mathscr{L}(\wp_l,m_k,\wp_{l+1})  \nonumber\\
    &+ \mathscr{L}(\wp_{l+1},m_k,\wp_{l+2}) +  \mathcal{V}_k^*(\wp_{l+2},m_k) \nonumber\\
    &\ldots \nonumber\\
    &\leq \sum_{\tau = l}^{T_f - 1} \mathscr{L}(\wp_{\tau},m_k,\wp_{\tau+1}) + \mathcal{V}_k^*(\wp_{T_f},m_k)\nonumber\\
    &= \sum_{\tau = l}^{T_f - 1} \mathscr{L}(\wp_{\tau},m_k,\wp_{\tau+1}) + \mathcal{V}_{k^{\prime}}^*(\wp_{T_f},m_k)\nonumber\\
    &= \mathcal{V}_{k^{\prime}}^*(\wp_l,m_k), \; k \leq l < T_f
    \label{eq:lower_bound_1}
\end{align}
where $\mathcal{V}_k^*(\wp_{T_f},m_k) = \mathcal{V}_{k^{\prime}}^*(\wp_{T_f},m_k) = [d(\wp_{T_f},\wp_{targ})]^2$ according to (\ref{eq:VF_m_k}). Because this holds for $l = T_f$ as well,  according to (\ref{eq:VF_m_k}), (\ref{eq:lower_bound_1}) can be rewritten by
\begin{equation}
	\mathcal{V}_k^*(\wp_l,m_k) \leq \mathcal{V}_{k^{\prime}}(\wp_l,m_k), \; k \leq l \leq T_f
	\label{eq:lower_bound_2}
\end{equation}

Next, consider the case of $k^{\prime} = k - 1$ and $0 \leq l \leq k^{\prime}$. Assume that there exists a sequence of optimal functional control laws $\mathcal{C}_{\tau}^*$, $\tau = l,\ldots,k^{\prime}$. By using these optimal functional control laws sequentially, a trajectory of robot PDFs $\{\wp_\tau\}_{\tau = l}^{k^{\prime}}$ is generated associated with $\{m_{\tau}\}_{\tau=l}^{k^\prime}$. 
Moreover, similarly, by using the optimal functional control law $\mathcal{C}_{k^\prime}^*$ recursively, a trajectory of robot PDFs $\{\wp_{\tau}\}_{\tau = k^{\prime}}^{T_f}$ is generated from  $\wp_{k^{\prime}}$ to $\wp_{T_f}$ associate with $m_k$. Thus, a trajectory of robot PDFs, $\{\wp_{\tau}\}_{\tau = l}^{T_f}$ is generated from $\wp_l$ to $\wp_{T_f}$. 

Again, according to (\ref{eq:VF_m_k}) and the Bellman equation,  $\mathcal{V}_k^*(\wp_l,\mathbf{M}_k)$ can be expressed by  
\begin{align}
\mathcal{V}_k^*(\wp_l,\mathbf{M}_k) &=  \sum_{\tau=l}^{k^{\prime}} \mathscr{L}(\wp_{\tau},m_{\tau},\mathcal{C}_{\tau}^*) + \mathcal{V}_k^*(\wp_k,m_k) \nonumber\\
&\leq   \sum_{\tau=l}^{k^{\prime}} \mathscr{L}(\wp_{\tau},m_{\tau},\mathcal{C}_{\tau}^*) + \mathcal{V}_{k^{\prime}}^*(\wp_k,m_k) \nonumber\\
&= \mathcal{V}_{k^{\prime}}^*(\wp_l,\mathbf{M}_k), \; 0 \leq l \leq k^{\prime}
\label{eq:lower_bound_3}
\end{align}
where the inequality is obtained by applying (\ref{eq:lower_bound_2}).

Merging (\ref{eq:lower_bound_2}) and (\ref{eq:lower_bound_3}), it is shown that for $k^{\prime} = k-1$, 
\begin{equation}
\mathcal{V}_k^*(\wp_l,\mathbf{M}_k) \leq \mathcal{V}_{k^\prime}^*(\wp_l,\mathbf{M}_k), \;  0 \leq l \leq T_f
\label{eq:lower_bound_4}
\end{equation} 

Finally, by recursively applying (\ref{eq:lower_bound_4}), the theorem is proved, such that
\begin{equation}
\mathcal{V}_k^*(\wp_l,\mathbf{M}_k) \leq \mathcal{V}_{k^\prime}^*(\wp_l,\mathbf{M}_k), \;0 \leq k^{\prime} \leq k \text{ and } 0 \leq l \leq T_f
\end{equation}
\end{IEEEproof}

\section{Upper bound of optimal value functional}
\label{Appedix: Upper_bound_of_optimal_value_functional}
\begin{IEEEproof}
	First, consider the case of $k+1 < T_f$. According to (\ref{eq:VF_WG}), the value functional $\mathcal{V}_k(\wp_{k+1},m_k,\tilde{\mathcal{C}}_k)$ associated to 
	$\tilde{\mathcal{C}}_k$, which is described in (\ref{eq:number_GC}) and (\ref{eq:weight_GC}), can be expressed by 
\begin{align}
	\mathcal{V}_k(\wp_{k+1},m_k,\tilde{\mathcal{C}}_k) &= \phi(\wp_{T_f},\wp_{targ}) + \sum_{\tau = k+1}^{T_f-1}\mathscr{L}(\wp_{\tau},m_k,\tilde{\mathcal{C}}) \nonumber\\
	&= \left[ d(\wp_{T_f},\wp_{targ}) \right]^2 \nonumber\\ 
	&+ \sum_{\tau =k+1}^{T_f -1} \left[\tilde{d} (\wp_{\tau},m_k,\tilde{\mathcal{C}}_k) \right]^2  \nonumber\\ &
	 + \sum_{\tau = k+1}^{T_f -1}   \langle   \wp_{\tau+1}, m_k \rangle_{\mathcal{W}} 	  
	\label{eq:VF_tilde_C}
	\end{align} 
According to (\ref{eq:def_WG_Metric}), the following inequality is obtained,
\begin{equation}
\left[ d(\wp_{T_f},\wp_{targ}) \right]^2 \leq 
\sum_{\imath=1}^{N_{k+1}}\sum_{j=1}^{N_{targ}} \left[ W_2(g^{\imath}_{T_f},g^j_{targ}) \right]^2 \tilde{\pi}_k(\imath,j)
\label{eq:d_sq_ineq}
\end{equation} 
By recursively applying (\ref{eq:number_GC}) and (\ref{eq:weight_GC}), the term, $\left[\tilde{d} (\wp_{\tau},m_k,\tilde{\mathcal{C}}_k) \right]^2$, $k+1 \leq \tau \leq T_f - 1$,  can be expressed as
\begin{align}
\left[\tilde{d} (\wp_{\tau},m_k,\tilde{\mathcal{C}}_k) \right]^2 &= \sum_{\imath=1}^{N_{k+1}}\sum_{\imath^{\prime}=1}^{N_{k+1}} [W_2(g^{\imath}_{\tau},g^{\imath^{\prime}}_{\tau+1})]^2 \pi_k(\imath,\imath^{\prime}) \nonumber\\
&= \sum_{\imath=1}^{N_{k+1}} \left[ W_2(g^{\imath}_{\tau},g^{\imath}_{\tau+1}) \right]^2 \omega_{\imath}^{k+1} \nonumber\\
&= \sum_{\imath=1}^{N_{k+1}} \sum_{j=1}^{N_{targ}} \left[ W_2(g^{\imath}_{\tau},g^{\imath}_{\tau+1}) \right]^2 \tilde{\pi}_k(\imath,j)
\label{eq:tilde_d_sq}
\end{align}

Substituting (\ref{eq:d_sq_ineq}) and (\ref{eq:tilde_d_sq}) into (\ref{eq:VF_tilde_C}), one can have
\begin{align}
\mathcal{V}_k(\wp_{k+1},m_k,\tilde{\mathcal{C}}_k) &\leq \sum_{\imath=1}^{N_{k+1}}\sum_{j=1}^{N_{targ}} \bigg\{ \left[ W_2(g^{\imath}_{T_f},g^j_{targ}) \right]^2 \nonumber\\
&+ \sum_{\tau =k+1}^{T_f -1} \left[ W_2(g^{\imath}_{\tau},g^{\imath}_{\tau+1}) \right]^2  \nonumber\\&
+ \sum_{\tau=k+1}^{T_f -1}  \langle   g^{\imath}_{\tau}, m_k \rangle_{\mathcal{W}} \bigg \} \tilde{\pi}_k(\imath,j)
\label{eq:upper_bound_1}
\end{align} 
Because (\ref{eq:upper_bound_1}) holds for any trajectories of Gaussian components from $g^{\imath}_{k+1}$ to $g^{j}_{targ}$, $\imath = 1,\ldots,N_{k+1}$ and $j = 1,\ldots,N_{targ}$, 
the following inequality can be obtained,
\begin{align}
\mathcal{V}_k(\wp_{k+1},m_k,\tilde{\mathcal{C}}_k) &\leq \sum_{\imath=1}^{N_{k+1}}\sum_{j=1}^{N_{targ}} \underset{\mathscr{Tr}^{\imath,j}_k}{\min}\bigg\{ \left[ W_2(g^{\imath}_{T_f},g^j_{targ}) \right]^2 \nonumber\\
&+ \sum_{\tau =k+1}^{T_f -1} \left[ W_2(g^{\imath}_{\tau},g^{\imath}_{\tau+1}) \right]^2  \nonumber\\&
+ \sum_{\tau=k+1}^{T_f -1}  \langle   g^{\imath}_{\tau}, m_k \rangle_{\mathcal{W}} \bigg \} \tilde{\pi}_k(\imath,j)\nonumber\\
&= \sum_{\imath=1}^{N_{k+1}} \sum_{j=1}^{N_{targ}} \tilde{\mathcal{L}}^{\imath,j}_k \tilde{\pi}_k(\imath,j) 
\label{eq:upper_bound_2}
\end{align} 

Next, consider the case of $k+1 = T_f$. According to (\ref{eq:d_sq_ineq}), the upper bound of $\mathcal{V}_k(\wp_{T_f},m_k,\tilde{\mathcal{C}}_k)$ is expressed by, 
\begin{align}
    \mathcal{V}_k(\wp_{T_f},m_k,\tilde{\mathcal{C}}_k) 
    &\leq  \sum_{\imath=1}^{N_{k+1}}\sum_{j=1}^{N_{targ}} \left[ W_2(g^{\imath}_{T_f},g^j_{targ}) \right]^2 \tilde{\pi}_k(\imath,j) \nonumber\\
    &= \sum_{\imath=1}^{N_{k+1}}\sum_{j=1}^{N_{targ}} \tilde{\mathcal{L}}^{\imath,j}_k \tilde{\pi}_k(\imath,j)
\end{align}

Finally, considering the definition of the optimal value functional, the theorem is proved, such that
\begin{align}
\mathcal{V}_k^*(\wp_{k+1},m_k) & \leq \mathcal{V}_k(\wp_{k+1},m_k,\tilde{\mathcal{C}}_k) \nonumber\\
& \leq  \sum_{\imath=1}^{N_{k+1}} \sum_{j=1}^{N_{targ}} \tilde{\mathcal{L}}^{\imath,j}_k  \tilde{\pi}_k(\imath,j) \nonumber\\
& = \tilde{\mathcal{V}}_k(\wp_{k+1},m_k)
\end{align}
\end{IEEEproof}

\section*{Acknowledgment}
This research was partially funded by National Science Foundation grant ECCS-1556900 and the Office 550 of Naval Research, Code 321.

\ifCLASSOPTIONcaptionsoff
  \newpage
\fi



\bibliographystyle{IEEEtran}
\bibliography{Ping_ADP_DOC_refs}
\end{document}